\documentclass[twocolumn, showpacs, preprintnumbers,%
amsmath,amssymb,prc]{revtex4}
\usepackage{graphicx}
\usepackage{bm}
\begin{document} 

\title{Irreducible tensor approach to spin observables in photo  
production\\
of mesons with arbitrary spin-parity $s^{\pi}$}
\author{G. Ramachandran}
 \affiliation{Indian Institute of Astrophysics, Koramangala,  
Bangalore, 560034, India}
\author{M. S. Vidya}
 \affiliation{V/5 RIE Campus, Regional Institute of Education, Mysore  
570006, India}
\author{J. Balasubramanyam}
 \affiliation{K. S. Institute of Technology,
Bangalore, 560062, India \\ 
Department of Physics, Bangalore University, Bangalore, 560056, India}

\date{\today}
\begin{abstract}
A theoretical formalism leading to elegant derivation of formulae for  
all spin observables is outlined for photo production of mesons with arbitrary  
spin parity $s^{\pi}$. The salient features of this formalism based on  
irreducible tensor techniques are  i) the number of independent irreducible tensor 
amplitudes is $4(2s+1)$, ii) a single compact formula is sufficient to  
express these amplitudes in terms of allowed electric and magnetic multipole  
amplitudes and iii) all the spin observables including beam analyzing powers as  
well as the differential cross section are expressible in terms of bilinear  
irreducible tensors of rank $0$ to $2(s+1)$. The relationship between the irreducible  
tensor amplitudes and the helicity amplitudes is elucidated in general and  
explicit expressions for the helicity amplitudes are given in terms of the  
irreducible tensor amplitudes in the particular cases of pseudoscalar and vector 
meson photo production. The connection between the irreducible tensor 
amplitudes introduced here and the well known CGLN amplitudes for  
photo production of pseudoscalar mesons is also established. \end{abstract}


\maketitle

\section{Introduction}
 
Photo production of heavy mesons on a nucleon is a topic of  
considerable current interest. Photo meson production has excited attention for more than  
five decades since the pion was discovered. In the context of developing 
a relativistic dispersion relation  approach to photo pion production, 
Chew, Goldberger, Low  and  Nambu  \cite{che}  have expressed the  reaction 
amplitude $\mathcal{F}$ in terms of  4 invariants viz., $ i   
\,{\boldsymbol 
\sigma} \cdot  \hat{{\boldsymbol \epsilon}},
\; ({\boldsymbol \sigma } \cdot \hat{{\boldsymbol q}})
({\boldsymbol \sigma} \cdot (\hat{{\boldsymbol k}}\times \hat{
{\boldsymbol \epsilon}})), \; i \,({\boldsymbol \sigma } \cdot 
\hat{{\boldsymbol k}}) (\hat{{\boldsymbol q}} \cdot \hat{{\boldsymbol 
\epsilon}})$  and  $  i\,({\boldsymbol \sigma} \cdot \hat{{\boldsymbol  q}})
(\hat{{\boldsymbol q}} \cdot \hat{{\boldsymbol \epsilon}}) $  and  
their respective coefficients  ${\mathcal F}_1, \;{\mathcal F}_2,   
\;{\mathcal F}_3$ and ${\mathcal F}_4$. The ${\boldsymbol \sigma }$ denote Pauli spin  
matrices of the nucleon,  the vectors ${\boldsymbol q}$ and ${\boldsymbol k}$ denote the meson 
and the photon momenta in the c.m. frame and $\hat{{\boldsymbol \epsilon}}$ denotes the photon  
polarization. The ${\mathcal F}_i , \, i=1,...,4$ are functions of  the c.m. 
energy $W$ at which the reaction takes place and the angle $\theta$  
between ${\boldsymbol q}$ and ${\boldsymbol k}$. 
Following the earlier work of Watson \cite{wat} 
and denoting the isospin index of the photo-produced pion by $\beta$, 
each of these amplitudes ${\mathcal F}_i$ were 
expressed in terms of three independent nucleon isospin combinations 
${\mathcal I}_{\beta}^{(+)},  \; {\mathcal I}_{\beta}^{(-)}$ and 
${\mathcal I}_{\beta}^{(0)}$ and 
their respective coefficients ${\mathcal F}_i^{(+)},  \;  {\mathcal  
F}_i^{(-)}$ and ${\mathcal F}_i^{(0)}$. Explicit 
formulae for ${\mathcal F}_i$ have been given \cite{che} in terms of  
the first two derivatives of Legendre polynomials in $\cos\theta$ and  
energy-dependent `magnetic' and 
`electric' multipole amplitudes denoted respectively by $M_{l\pm}$ and  
$E_{l\pm}$, where the suffix $\pm$ indicates that the total angular momentum 
$j = l\pm {\textstyle{\frac{1}{2}}}$,
if $l$ denotes the orbital angular momentum of the emitted meson. This  
traditional formalism has been 
reviewed along with the extension \cite{phd} to 
electro production by  Berends, Donnachie and 
Weaver \cite{ber} in a set  of  three papers. The connection with the  
helicity formalism \cite{jac} for photo production as well as formulae for the 
differential cross-section and spin observables in terms of the CGLN  
amplitudes are also found in \cite{ber} along 
with numerical values for the amplitudes up to 500 MeV. The dominance  
of the first resonance viz., $\Delta_{33}(1232)$ is quite 
conspicuous in this energy region and as such attention has lately  
been focussed \cite{burk,buch} on details such as the quadrupole deformation of  
this resonance and ratio of $E_{1+}$ over 
$M_{1+}$. In view of the necessity 
of neutron multipoles \cite{moore} to determine ${\mathcal F}^{(+)},   
\;  {\mathcal F}^{(-)}$ and ${\mathcal F}^{(0)}$ and their importance  to decide questions of 
time reversal violation \cite{donna} or the possible existence of  
isotensor term \cite{san}, a careful discussion of target asymmetry  and effective neutron  
polarization was presented \cite{rsk} as also detailed theoretical analyses to extract the  
neutron multipoles more precisely from experiments on hydrogen isotopes \cite{rsk2}.
Photo pion production has been studied  extensively and reviewed by  several groups \cite{rev}. 
The current database as well as numerical values for the electric 
and magnetic multipole amplitudes derived from 
accumulated data can be found in the Center for Nuclear 
Studies(CNS) website \cite{cns}. Observables have  been defined 
 in terms of helicity and transversity amplitudes \cite{bark}. 
The helicity amplitudes \cite{walker} can also be constructed from the  
multipole amplitudes \cite{cns} using relations found in \cite{arndt}.
As photon energy is increased, the nucleon resonances 
$P_{11}(1440), D_{13}(1520), S_{11}(1535)$ and $S_{11}(1650)$ start 
contributing along with the first and higher $\Delta$ resonances  
\cite{eid}.

In contrast to the isovector pion, the $\eta$
meson is isoscalar. Consequently, the $\Delta$ resonances do not  
contribute to photo production of $\eta$ and as such it is ideally suited to study  
the nucleon resonances. The enigmatic \cite{burk,rop} Roper resonance  
$P_{11}(1440)$ lies below $\eta$ production threshold, so that only
its high energy tail can contribute. Close to threshold, the  
contributions of $P_{11}(1440)$ and $D_{13}(1520)$ with larger $l$ values are 
suppressed compared to the $S_{11}(1535)$ and $S_{11}(1650)$. 
However, the decay pattern of the first and second $S$ wave nucleon  
resonances were found to be quite different and are not easily explained in terms  
of nucleon models. In fact, a variety of nucleon models predict a much richer nucleon 
excitation spectrum than what has been identified in $\pi N$  
scattering. Considerable interest has, therefore, been evinced in looking for the 
so-called `missing resonances' \cite{mis}. 
Reference may be made to theoretical studies \cite{ELA} based on  
Effective Lagrangian Approach. Excellent reviews \cite{kru} exist on  
these developments and the problem of `missing resonances' in photo  
production reactions.  As $\eta$ is also a pseudoscalar like the pion, the spin structure of  
the amplitudes for photo production of $\eta$ is  of the same form
as for photo pion production and the corresponding
multipole amplitudes can also be found in \cite{cns}.

The associated strangeness photo production has been studied  \cite{hyp} 
over the past several decades employing different approaches. In contrast to photo
production of pions and eta, kaon photo production can involve nucleon as well as hyperon 
resonances like  $ S_{01}(1405), \, S_{01}(1670),\, 
P_{01}(1810)$ and  some models include nucleon resonances with spin greater than $3/2$. 
It was remarked \cite{mart}  recently that   $ S_{11}(1650), \, P_{13}(1720),
D_{13}(1700), \, D_{13}(2080), \, F_{15}(1680)$ and $ F_{15}(2000)$ are 
required to fit one set of experimental data,  while   
 $P_{13}(1900), \, D_{13}(2080), \, D_{15}(1675), \, 
F_{15}(1680)$ and $F_{17}(1990)$ resonances are required to fit another set. 
Although  both   the sets do not exhibit 
the need for $P_{11}(1710)$ and are agreed that the second peak 
in the cross-sections at $W \sim 1900$ MeV originates  
from the  $D_{13}(2080)$ resonance \cite{eid} whose mass lies between $1911$ MeV  
and $1936$ MeV, fitting to all data simultaneously changes the conclusion and 
results in a model that is inconsistent  to all data sets.
It is clear that the delta resonances can also contribute in addition to the nucleon 
resonances in the case of  $\gamma p \to \Sigma  K$. The beam analyzing 
powers \cite{zeg} and beam-recoil observables \cite{brad} have also been
measured very recently.  In hyperon photo production, 
the hyperons have spin parity  ${\textstyle\frac{1}{2}}^{+}$ just like the nucleons, 
while kaons are pseudoscalar like the pion. Consequently the spin  
structure of the amplitudes for hyperon and pion photo production are alike \cite{fasa}, although 
the isospin considerations are different as with  the case of photo  
production of $\eta$.

As photon  energy is increased further, the thresholds are 
reached for $\rho, \eta', \omega$ and $\varphi$ of 
which $\eta'$ is pseudoscalar and $\rho, \omega, 
\varphi$ are vector mesons. Experimental studies on photo production 
of $\rho$ and $\omega$ date back to the early sixties and it has been  
observed that the reaction is governed by diffraction or pomeron  
exchange at  higher energies. The inherent advantages in  employing linearly  
polarized photons  \cite{cooper} have  been noted and  a detailed formalism to  
analyze production of vector mesons with polarized photons has also been presented
 \cite{schil},   employing twelve  complex helicity amplitudes. 
 A historical perspective and 
a detailed account of the theoretical ideas that motivated early studies 
on vector meson photo production may be found in \cite{bauer}. 
In the case of photo production of a vector 
meson,  the meson polarization  itself is an interesting observable which has been 
studied experimentally in recent years \cite{wu} 
and analyzed using the formalism of  \cite{schil}. 
All the polarization observables were expressed as bilinear 
\cite{tab961} products of the twelve helicity amplitudes and their 
behavior  near threshold \cite{tab962} has  been examined.     
It was shown \cite{tab98} that a  two meson  decay of a 
$\rho$ or $\varphi$  does not determine its vector polarization.  
More recently \cite{a01}, it was shown that the three meson decay of 
$\omega$ also does not determine its vector polarization. On the other 
hand, the $\pi_{0}\gamma$ decay mode of $\vec{\omega}$
with a smaller branching ratio of 8.92\% may be utilized to determine
its vecor as well as tensor polarization \cite{a02}.
The isovector $\rho$ has a large width of 150.2 MeV, as compared to  
the  widths of  8.44 MeV and 4.458 MeV of the isoscalar $\omega$ 
and $\varphi$ respectively and of 1.18 $\times 10^{-3}$ MeV and 0.2 
MeV respectively of $\eta$ and $\eta'$. 
There are hardly any  experimentally known resonances  which decay 
into $N\omega$,  except $N(1710)$ which has a branching ratio \cite{eid} of 
$\sim 13\%$. However, there are theoretical expectations 
that missing resonances may couple more strongly or even exclusively  
to the  $ N\omega$ channel in comparison with the $ N\pi$ channel. 
Hence  
several theoretical  studies have been carried out \cite{n1} on the 
contribution of nucleon resonances in $\omega$ 
photo production.  
Another interesting aspect of vector meson photo production is 
the $\phi/\omega$ ratio \cite{n3} in the context of the violation of  
OZI rule \cite{n4}. This energy region above the threshold for photo production of 
$\eta, \rho, \omega, \eta'$ and $\varphi $ is under intense study at  
present with the advent of the new generation 
of electron accelerators like CEBEF at JLab, ELSA at Bonn, ESRF at  
Grenoble, MAMI at Mainz and  Spring8 at Osaka 
which are equipped with tagged photon facilities where the incident  
photon energies are known event by event within the resolution of the tagging  
detector.  While photon beams are produced by bremsstrahlung at 
JLab(CLAS), ELSA(SAPHIR) and MAMI, laser back scattering is employed  
at BNL(LEGS), ESRF(GRAAL) and Spring8(LEPS). Availability of linearly and 
circularly polarized photons as well as polarized targets enable  
measurements  of beam and target analyzing powers, in addition to the differential  
cross-section. Such measurements are useful to determine resonance parameters. 
In future one may look  
forward to photo production of higher spin mesons \cite{eid}, like $f_2(1270)$ or 
$a_2(1320)$ or $f_2'(1525)$ with spin-parity $2^+$,  $\pi_2(1670)$ 
with spin-parity $2^-$,  $\omega_3(1670)$ or $\rho_3(1690)$ with  
spin-parity $3^-$ and  
$a_{4}(2040)$ or $f_{4}(2050)$ with spin-parity $4^+$. 
An exotic baryon state with mass $M=1555\pm10$ MeV 
and strangeness $+1$ was observed \cite{a03} in photo production from the proton.  
More recently, a theoretical analysis has been carried out on meson  
spectra in a generalized constituent quark model \cite{n2} which pays 
attention also to glueballs, hybrids or multiquark states. 
In view of  
these developments, it is felt that a supportive theoretical   
formalism for the spin structure of the amplitudes and their  expansion  
in terms of `electric' and `magnetic' multipoles is needed to 
analyze  measurements of spin observables in photo production of  
mesons with arbitrary spin-parity $s^{\pi}$.

The purpose of the present paper is to answer this need. 
Employing irreducible tensor operator techniques \cite{gr1} 
we derive formulae for all  spin observables, including  beam  
analyzing powers, associated with photo production of mesons with isospin $I_m$ 
and  arbitrary spin-parity $s^\pi$ after outlining the basic  
theoretical formalism in the next section. 

\section{Theoretical formalism}

Let ${\boldsymbol q}$ and ${\boldsymbol k}$ denote the meson and the 
photon momenta in c.m. frame for photo production of arbitrary spin  
mesons at  c.m. energy $W$. If $\hat{\boldsymbol u}_x$, $\hat{\boldsymbol u}_y$  
denote unit vectors of a right 
handed Cartesian coordinate system with ${\boldsymbol k}$ along the  
z-axis, the left and right circular polarized states of the photon may be  
defined, following Rose \cite{ros}, through 
\begin{equation}
\label{uv}
\hat{\boldsymbol u}_{\mu}= \frac{1}{\sqrt{2}}(\hat{\boldsymbol  
u}_x+i\mu  \hat{\boldsymbol u}_y)=-\mu \hat{\boldsymbol \xi}_{\mu},
\; \mu = \pm 1.
\end{equation}

When the reaction is initiated by photons polarized in a state 
$\hat{\boldsymbol u}_{\mu}$, the differential cross-section in c.m. 
frame may be written as 
\begin{equation}
\label{gde}
\frac{d\sigma(\mu)}{d\Omega} = \frac{q}{k}\sum \sum_{av}|\langle f|
{\mathcal F(\mu)}|i\rangle|^2,
\end{equation}
where $|i\rangle$ and $|f\rangle$ denote respectively the initial and  
final  hadron spin states, ${\boldsymbol q}$ and ${\boldsymbol k}$  have  
polar coordinates $(q,\theta, \varphi)$ and $(k,0,0)$ respectively and  
$\sum$ denotes summation with respect to final and $\sum_{av}$ the average with  
respect to initial hadron spin states. 

When a meson with spin $s$ is photo-produced on a nucleon, the channel  
spin $s_f$ in the final state could assume either of the values 
$|s-{\textstyle{\frac{1}{2}}}|$ and $(s+{\textstyle{\frac{1}{2}}})$  
and the transition to the final hadron system with spin $s_f$ takes place  
from the  initial state of the hadron with spin $s_i={\textstyle{\frac{1}{2}}}$.  
We may, therefore write the reaction amplitude ${\mathcal F}(\mu)$ in the form
\begin{equation}
\label{ra}
{\mathcal F}(\mu) = \sum_{s_f=|s-{\textstyle{\frac{1}{2}}}|}^{(s+
{\textstyle{\frac{1}{2}}})}\sum_{\lambda=
|s_f-{\textstyle{\frac{1}{2}}}|}^{(s_f+{\textstyle{\frac{1}{2}}})}
\big(S^{\lambda}(s_f,{\textstyle{\frac{1}{2}}})
\cdot  {\mathcal F}^{\lambda}(s_f, \mu)\big)
\end{equation}
in terms of irreducible tensor operators  
$S_{m_{\lambda}}^{\lambda}(s_f,s_i)$
of rank $\lambda$ in hadron spin space defined in \cite{gr1}. To  
identify the irreducible tensor amplitudes  ${\mathcal  
F}_{m_{\lambda}}^{\lambda}(s_f,\mu)$,
we may evaluate  $\langle f|{\mathcal F}(\mu)|i\rangle$ by writing it  
explicitly as 
\begin{eqnarray}
\label{ita}
&&\langle(s {\textstyle{\frac{1}{2}}})s_fm_f;{\boldsymbol q}\,
|{\mathcal F}|\,
{\boldsymbol k} \hat{\boldsymbol  
u}_{\mu};{\textstyle{\frac{1}{2}}}m_i\rangle =
4\pi (2\pi)^{{\textstyle{\frac{1}{2}}}} \nonumber \\ && \times 
\sum_{l=0}^{\infty}(-i)^
l \sum_{L=1}^{\infty} i^L [L] 
\sum_{j=L-{\textstyle{\frac{1}{2}}}}^{L+{\textstyle{\frac{1}{2}}}} 
C(L{\textstyle{\frac{1}{2}}}j;\mu m_i m) \nonumber \\ && \times 
C(l s_f j;m_lm_fm) Y_{lm_l}(\theta, \varphi)(i\mu)^{f_{+}(L,l)}
 {\mathcal F}_{ls_f;L}^j \ ,
\end{eqnarray}
using the standard multipole expansion  \cite{ros} for the photon in  
the initial  state and partial wave expansion for the meson in the final state. We  
denote
$\sqrt{(2L+1)}$ by $[L]$. The rest of the notations 
follow Rose \cite{ros}. Using
\begin{equation}
f_{\pm}(L,l)= \frac{1}{2}[1\pm \pi(-1)^{L-l}]
\end{equation}
and parity conservation, we may express
\begin{eqnarray}
\label{pwmpa}
 {\mathcal F}_{ls_f;L}^j  &\equiv& \langle  
(l(s{\textstyle{\frac{1}{2}}})s_f)j\|
{\mathcal F}\|(L {\textstyle{\frac{1}{2}}})j \rangle \nonumber \\
&=&  {\mathcal M}^j_{ls_f;L} f_-(L,l)+  {\mathcal E}^j_{ls_f;L}  
f_+(L,l)
\end{eqnarray}
in terms of `magnetic' and `electric' multipole amplitudes denoted by 
$ {\mathcal M}^j_{ls_f;L} $ and  $ {\mathcal E}^j_{ls_f;L} $  
respectively. We rewrite the two Clebsch-Gordan coefficients in Eq. \eqref{ita} as 
\begin{eqnarray}
&&C(L{\textstyle{\frac{1}{2}}}j;\mu m_i m) C(l s_f j;m_lm_fm) =  
\sum_{\lambda}
W(L{\textstyle{\frac{1}{2}}}ls_f;j\lambda)\nonumber \\
&&\times [j]^2[s_f]^{-1}(-1)^{L+l+{\textstyle{\frac{1}{2}}}-j}
(-1)^{m_{\lambda}}
C(lL\lambda ;m_l -\mu m_{\lambda})\nonumber \\ &&\times
(-1)^{\mu}C({\textstyle{\frac{1}{2}}} \lambda s_f ; m_i  
-m_{\lambda}m_f)[\lambda]
\end{eqnarray}
and replace
\begin{equation}
\label{so}
C({\textstyle{\frac{1}{2}}} \lambda s_f ; m_i  
-m_{\lambda}m_f)[\lambda]=
\langle s_f m_f |  
S^{\lambda}_{-m_{\lambda}}(s_f,{\textstyle{\frac{1}{2}}})
|{\textstyle{\frac{1}{2}}} m_i\rangle .
\end{equation}

Comparing the resulting expression with $\langle f|{\mathcal  F}(\mu)|i\rangle$ and 
using Eq. \eqref{ra} we obtain the elegant and compact formula for the  
irreducible tensor amplitudes
\begin{eqnarray}
\label{cf}
{\mathcal F}^{\lambda}_{m_{\lambda}}(s_f,\mu) &=& 
4\pi (2\pi)^{{\textstyle{\frac{1}{2}}}} 
\sum_{l=0}^{\infty} \sum_{L=1}^{\infty} (i)^{L-l}
\sum_{j}(-1)^{L+l+{\textstyle{\frac{1}{2}}}-j}\nonumber \\
&& \times[j]^2 
[L][s_f]^{-1} W(L{\textstyle{\frac{1}{2}}}ls_f;j\lambda) \nonumber  
\\&&
\times (i\mu)^{f_+(L,l)} {\mathcal F}_{ls_f;L}^j   
(-1)^{\mu}\nonumber \\&&
\times C(lL\lambda ;m_l -\mu m_{\lambda}) Y_{lm_l}(\theta, \varphi),
\end{eqnarray}
in terms of partial wave multipole amplitudes $ {\mathcal  F}_{ls_f;L}^j $ given 
by Eq. \eqref{pwmpa} for photo production of mesons with arbitrary  
spin-parity,  $s^{\pi}$. 

Isospin considerations lead to 
\begin{eqnarray}
\label{ispa}
 {\mathcal F}_{ls_f;L}^j  &=& \sum_{I=|I_m-{\textstyle{\frac{1}{2}}}|}
^{(I_m+{\textstyle{\frac{1}{2}}})} 
C({\textstyle{\frac{1}{2}}} I_m I ;  \nu_f  \nu_m \nu) \nonumber \\ &  
& \times \sum_{I_{\gamma}=0}^{1}
C({\textstyle{\frac{1}{2}}} I_{\gamma}I;\nu_i  0 \nu)
{\mathcal F}_{ls_f;L}^{I_{\gamma}I j}\ ,
\end{eqnarray}
where $\nu_i , \nu_f , \nu_m$ denote respectively the isospin  
projection quantum numbers of the nucleon in the initial and final states and the meson  
which is  photo produced. Dropping the indices $j, l, s_f, L$ common to both  
sides of  Eq. \eqref{ispa},  the  ${\mathcal F}^{I_{\gamma}I }$  thus defined  
for photo pion production may readily be related to the ${\mathcal F}^{(\pm)}$, 
${\mathcal F}^{(0)}$ of CGLN \cite{che} as shown  in the appendix A. 
The isospin indices $I_{\gamma}, I$ may  
also be attached using  Eq. \eqref{ispa} to the `magnetic' and   `electric' multipole
amplitudes defined by Eq. \eqref{pwmpa}. If one is looking for the  
possible existence  of an isotensor component \cite{san}, the summation over $I_{\gamma}$  
may readily  be extended in Eq. \eqref{ispa} to include $I_{\gamma}=2$. The  
advantage in having the 
index $I$ along with $j$ in this formalism is  
that it facilitates ready identification with the isospin and spin quantum  
numbers of the  resonances which contribute in the intermediate state to photo production
of mesons with arbitrary spin-parity $s^\pi$.

It may be noted that the spherical harmonics $Y_{lm_l}$ in  Eq. \eqref{cf}
contain the azimuthal angle $\varphi$ along with the polar angle $\theta$ of 
the momentum ${\boldsymbol q}$ of the meson.  This facilitates the analysis 
of experiments on photo meson production employing linearly polarized photons where the 
state of linear polarization of the beam is  chosen to be along the x-axis.

\subsection{Irreducible tensor amplitudes in the Madison Frame}
 In discussing 
 photo meson production with a two body  final state, it is quite 
often convenient to chose the reaction plane containing ${\boldsymbol k}$  and
${\boldsymbol q}$  as the z-x plane in which case   the azimuthal angle $\varphi = 0$.
In fact this choice of the Cartesian  coordinate system 
has generally been recommended by the Madison Convention 
\cite{mad}.  We may therefore  refer  to  this Cartesian  coordinate system
as the  Madison Frame $(MF)$. The  irreducible tensor amplitudes  
${\mathcal F}^{\lambda}_{m_{\lambda}}(s_f, \mu)$
in $MF$ are then given by Eq. \eqref{cf} with $\varphi = 0$ , in which case they 
satisfy 
\begin{equation}
\label{sr}
{\mathcal F}^{\lambda}_{-m_{\lambda}}(s_f, -\mu)= \pi 
(-1)^{\lambda - m_{\lambda}}
{\mathcal F}^{\lambda}_{m_{\lambda}}(s_f, \mu).
\end{equation}

In view of the above, the total number
\begin{equation}
N_{tot}= \sum_{\mu=-1,1}\sum_{s_f=|s-{\textstyle{\frac{1}{2}}}|}^{(s+
{\textstyle{\frac{1}{2}}})}\sum_{\lambda=
|s_f-{\textstyle{\frac{1}{2}}}|}^{(s_f+{\textstyle{\frac{1}{2}}})}
(2\lambda+1) =8(2s+1),
\end{equation}
of  irreducible   tensor amplitudes reduces to $4(2s+1)$  independent irreducible  
tensor amplitudes. This number is exactly in agreement with four \cite{che}  
for $s=0$ and twelve \cite{schil} for $s=1$  arrived at by using different  
arguments in those particular cases. 
In the case of electro production \cite{pramana}, the longitudinal  
polarization state with $\mu = 0$  contributes  additional  $2(2s+1)$ independent 
amplitudes and consequently the total number of independent amplitudes 
for electro production turns out  to be $6(2s+1)$. It may be noted that $ {\mathcal  
M}^j_{ls_f;L}, {\mathcal E}^j_{ls_f;L} $ and  ${\mathcal F}_{ls_f;L}^j$ in  
\cite{pramana} are 
$4\pi^{{\textstyle\frac{3}{2}}}i^{l+L}   
(-1)^{L+{{\scriptstyle\frac{1}{2}}}-j}[L][j]^2[s_f][s]^{-1}$ 
times the $ {\mathcal M}^j_{ls_f;L},  {\mathcal E}^j_{ls_f;L} $ 
and $ {\mathcal F}_{ls_f;L}^j $ given by Eq. \eqref{pwmpa} and that 
the irreducible tensor amplitudes ${\mathcal  
F}^{\lambda}_{m_{\lambda}}(n,\mu), n=0,1$ 
introduced in \cite{pramana} may also be expressed in terms of the 
${\mathcal F}^{\lambda}_{m_{\lambda}}(s_f, \mu)$ through
\begin{eqnarray}
{\mathcal F}^{\lambda}_{m_{\lambda}}(n ,\mu) &=&
\frac{1}{\sqrt{2}}[n][s]^{-1}\sum_{s_f}(2s_f+1)
\nonumber \\
&\times&  W(\lambda  {\textstyle{\frac{1}{2}}}s{\textstyle{\frac{1}{2}}};s_fn)
{\mathcal F}^{\lambda}_{m_{\lambda}}(s_f, \mu),
\end{eqnarray}
or conversely
\begin{equation}
{\mathcal F}^{\lambda}_{m_{\lambda}}(s_f,\mu) =
\sqrt{2}[s][s_f]^{-1}\sum_{n}W(\lambda 
{\textstyle{\frac{1}{2}}}s{\textstyle{\frac{1}{2}}};s_fn)
{\mathcal F}^{\lambda}_{m_{\lambda}}(n, \mu).
\end{equation}

It may be noted that the irreducible tensor operators $S^n_{m_n}
({\textstyle{\frac{1}{2}}},{\textstyle{\frac{1}{2}}})$ of rank $n$ in 
\cite{pramana} are identified as 
\begin{eqnarray}
S^0_0({\textstyle{\frac{1}{2}}},{\textstyle{\frac{1}{2}}})  &=& 1  \nonumber \\
S^1_{\pm 1}({\textstyle{\frac{1}{2}}},{\textstyle{\frac{1}{2}}})  &=&
\mp \frac{1}{\sqrt{2}} (\sigma_x \pm i \sigma_y) \\ 
S^1_0({\textstyle{\frac{1}{2}}},{\textstyle{\frac{1}{2}}})  &=& \sigma_z,
\nonumber 
\end{eqnarray}
in terms of Pauli matrices. It may  perhaps be mentioned that in the case of 
kaon photo production, the rows and columns of these matrices are to be labeled 
by the spin states of the hyperon and nucleon respectively.  In the  particular  
case of photo production of pseudoscalar mesons,
the connection between our amplitudes given by Eq. \eqref{cf} with  
$\phi =0$  in the Madison Frame \cite{mad} and 
those of CGLN is established in Appendix A.

\subsection{Irreducible tensor amplitudes in the Transverse Frame}
The Transverse Frame $(TF)$ may be defined as the right handed Cartesian  
coordinate system with the z-axis chosen along ${\boldsymbol k}\times
{\boldsymbol q}$, i.e., transverse to the reaction plane and with the 
x-axis chosen along ${\boldsymbol k}$.  The explicit form for 
 $\langle f|{\mathcal F}(\mu)|i\rangle$,  in this frame given by
\begin{eqnarray}
&&\langle(s {\textstyle{\frac{1}{2}}})s_fm_f;{\boldsymbol q}\,
|{\mathcal F}|\,
{\boldsymbol k} \hat{\boldsymbol  
u}_{\mu};{\textstyle{\frac{1}{2}}}m_i\rangle =
4\pi (2\pi)^{{\textstyle{\frac{1}{2}}}} \nonumber \\ && \times 
\sum_{l=0}^{\infty}(-i)^
l \sum_{L=1}^{\infty} i^L [L] 
\sum_{j=L-{\textstyle{\frac{1}{2}}}}^{L+{\textstyle{\frac{1}{2}}}} 
(i\mu)^{f_{+}(L,l)}
 {\mathcal F}_{ls_f;L}^j  \nonumber \\ && \times 
\sum_{M=-L}^{L}\nonumber  
C(L{\textstyle{\frac{1}{2}}}j;M m_i m) 
C(l s_f j;m_lm_fm) \nonumber \\ && \times 
D^{L}_{M\mu}(0,{\textstyle{\frac{\pi}{2}}},0)Y_{lm_l}({\textstyle{\frac 
{\pi}{2}}},\theta),
\end{eqnarray}
instead of \eqref{ita}, so that 
the irreducible tensor amplitudes in $TF$ are given by
\begin{eqnarray}
\label{ittf}
{\mathcal F}^{\lambda}_{m_{\lambda}}(s_f,\mu)_{TF} &=& 
4\pi (2\pi)^{{\textstyle{\frac{1}{2}}}} 
\sum_{l=0}^{\infty} \sum_{L=1}^{\infty} (i)^{L-l}
\sum_{j}(-1)^{L+l+{\textstyle{\frac{1}{2}}}-j}\nonumber \\
&& \times[j]^2 
[L][s_f]^{-1} W(L{\textstyle{\frac{1}{2}}}ls_f;j\lambda) \nonumber  
\\&&
\times  {\mathcal F}_{ls_f;L}^j   
{\mathcal A}^{\lambda}_{m_{\lambda}}(\mu,\theta),
\end{eqnarray}
where
\begin{eqnarray}
{\mathcal A}^{\lambda}_{m_{\lambda}}(\mu,\theta) &=& 
(i\mu)^{f_+(L,l)}
\sum_{M}(-1)^{M}
C(lL\lambda ;m_l -M m_{\lambda})\nonumber \\ && \times 
 Y_{lm_l}({\textstyle{\frac{\pi}{2}}},\theta)
d_{M\mu}^{L}({\textstyle{\frac{\pi}{2}}}),
\end{eqnarray}
which now defines the angular distribution of the 
meson as a function of  $\theta$ in the $TF$.

The amplitudes \eqref{ittf} are related to 
${\mathcal F}^{\lambda}_{m_{\lambda}}(s_f,\mu)$ in the  
Madison frame  through 
\begin{equation}
\label{tfmf}
{\mathcal F}^{\lambda}_{m_{\lambda}}(s_f,\mu)_{TF}=
\sum_{m_{\lambda}^{\prime}=-\lambda}^{+\lambda}
D^{\lambda}_{m_{\lambda}^{\prime}m_{\lambda}}({\textstyle{\frac{\pi}{2 
}}},{\textstyle{\frac{\pi}{2}}},\pi)
{\mathcal F}^{\lambda}_{m_{\lambda^{\prime}}}(s_f,\mu).
\end{equation}

In fact, the irreducible tensor amplitudes ${\mathcal F}^{\lambda}_{m_{\lambda}}
(s_f,\mu)_{AF}$ in any frame ($AF$) may be expressed in terms of the 
${\mathcal F}^{\lambda}_{m_{\lambda}}(s_f,\mu)_{GF}$ in any given
frame ($GF$ ) through

\begin{equation}
\label{afgf}
{\mathcal F}^{\lambda}_{m_{\lambda}}(s_f,\mu)_{AF} =
\sum_{m_{\lambda}'} D_{m_{\lambda}' m_{\lambda}}^{\lambda}
(\alpha, \beta, \gamma) {\mathcal F}^{\lambda}_{m_{\lambda}'}(s_f,\mu)_{GF}
\end{equation}
if $(\alpha, \beta, \gamma)$ denote the Euler angles \cite{ros} which
characterize  rotation connecting $AF$ from  $GF$. 
The relationship between the irreducible tensor 
amplitudes and helicity amplitudes is elucidated in Appendix B.

It is sometimes convenient to chose a frame whose z-axis is along 
${\boldsymbol q}$, y-axis is along $({\boldsymbol k} \times {\boldsymbol q})$
and x-axis is along $({\boldsymbol k} \times {\boldsymbol q})\times {\boldsymbol q}$,  
in which case $\alpha = 0, \beta = \theta$ and $\gamma = 0$, when $GF$ is  identified 
with $MF$. The spin states $|sm_s\rangle$  of the meson defined with respect to this frame remain the 
same,  when we make a Lorentz transformation along ${\boldsymbol q}$ to reach the 
meson rest frame. This Lorentz transformation is needed en-route to 
Gottfried-Jackson frame \cite{gott}.

We will express the differential cross-section and all spin observables in the next 
section in terms of bilinear irreducible tensors $B^{\Lambda}_{m_{\Lambda}}$ of 
rank $ \Lambda$ constructed out of the 
${\mathcal F}^{\lambda}_{m_{\lambda}}(s_f,\mu)$. It may be noted that 
the $B^{\Lambda}_{m_{\Lambda}}$ which are related to observables measurable 
experimentally which may conveniently be obtained in any required  frame by  choosing 
irreducible tensor amplitudes in the appropriate frame.

\section{Differential cross-section and spin observables}
The unpolarized differential cross-section, in c.m. frame, is given by
\begin{equation}
\label{updc}
\frac{d\sigma_0}{d\Omega}=\frac{1}{2}\sum_{\mu=-1,1}
\frac{d\sigma(\mu)}{d\Omega},
\end{equation}
where $d\sigma(\mu)/d\Omega$ defined by Eq. \eqref{gde} is readily  
given, on using 
Eqs. \eqref{ra} and \eqref{so} by
\begin{equation}
\label{updcf}
\frac{d\sigma(\mu)}{d\Omega} = \frac{q}{2k}\sum_{s_f}(2s_f+1)
\sum_{m_{\lambda}}
|{\mathcal F}^{\lambda}_{m_{\lambda}}(s_f,\mu)|^2.
\end{equation}

\subsection{\label{3a}Beam analyzing powers}
The beam analyzing power $\Sigma_3$ with respect to left and right  
circular  polarized states is readily given by
\begin{equation}
\frac{d\sigma_0}{d\Omega}\Sigma_3 = \frac{d\sigma(+1)}{d\Omega}-
\frac{d\sigma(-1)}{d\Omega},
\end{equation}
where $d\sigma_0/d\Omega$ and $d\sigma(\mu)/d\Omega$ for $\mu = 
\pm1$ are known
from Eqs. \eqref{updc} and \eqref{updcf}.
It is clear from Eq. \eqref{uv} that
\begin{equation}
\hat{\boldsymbol u}_x= \frac{1}{\sqrt{2}}(\hat{\boldsymbol u}_{+1}+ 
\hat{\boldsymbol u}_{-1}) \; ; \; 
\hat{\boldsymbol u}_y= \frac{-i}{\sqrt{2}}(\hat{\boldsymbol u}_{+1}- 
\hat{\boldsymbol u}_{-1}),
\end{equation}
so that photons linearly polarized along an azimuthal angle $\alpha$  
are represented by
\begin{eqnarray}
\hat{\boldsymbol u}_{\alpha}&=& \hat{\boldsymbol u}_x \cos\alpha+
\hat{\boldsymbol u}_y \sin\alpha \nonumber \\
&=& \frac{1}{\sqrt{2}}[\hat{\boldsymbol u}_{+1}e^{-i\alpha}+
\hat{\boldsymbol u}_{-1}e^{i\alpha}].
\end{eqnarray}

The differential cross-section, when the beam is linearly polarized  
along $\alpha$ is then given by
\begin{eqnarray}
\label{dcbp}
\frac{d\sigma(\alpha)}{d\Omega}&=& \frac{q}{4k}\sum_{s_f}(2s_f+1)
\sum_{\lambda}
\sum_{m_{\lambda}}\nonumber \\ &\times&
|{\mathcal F}^{\lambda}_{m_{\lambda}}(s_f, 1)e^{-i\alpha}+
{\mathcal F}^{\lambda}_{m_{\lambda}}(s_f, -1)e^{i\alpha}|^2
\end{eqnarray}
using which we may readily define the beam analyzing powers $\Sigma_1$  
and $\Sigma_2$, with respect to linearly polarized states, through 
\begin{eqnarray}
\frac{d\sigma_0}{d\Omega}\Sigma_1 &=& \frac{d\sigma(\alpha=0)}
{d\Omega}-
\frac{d\sigma(\alpha=\pi/2)}{d\Omega} \\
&=& \frac{q}{k}\sum_{s_f}[s_f]^2\sum_{\lambda}
\sum_{m_{\lambda}}\nonumber \\ &\times& 
\Re[{\mathcal F}^{\lambda}_{m_{\lambda}}(s_f, 1)
{\mathcal F}^{\lambda}_{m_{\lambda}}(s_f, -1)^*],
\end{eqnarray}
\begin{eqnarray}
\frac{d\sigma_0}{d\Omega}\Sigma_2 &=& \frac{d\sigma(\alpha=\pi/4)}
{d\Omega}-\frac{d\sigma(\alpha=3\pi/4)}{d\Omega} \\
&=& \frac{q}{k}\sum_{s_f}[s_f]^2\sum_{\lambda}
\sum_{m_{\lambda}}\nonumber \\ &\times&
\Im[{\mathcal F}^{\lambda}_{m_{\lambda}}(s_f, 1)
{\mathcal F}^{\lambda}_{m_{\lambda}}(s_f, -1)^*],
\end{eqnarray}
where ${\mathcal F}^{\lambda}_{m_{\lambda}}(s_f, \mu)^*$ denotes the 
complex  conjugate of Eq. \eqref{cf}.

We may note that the analyzing powers $\Sigma_1, \; \Sigma_2, \;  
\Sigma_3$ correspond 
respectively to the well known Stokes parameters $s_1, s_2, s_3$ in  
terms of  which the state of polarization of the beam may be described by the  
density  matrix
\begin{equation}
\rho^{\gamma} = \frac{1}{2}[1+\sigma_1^{\gamma}s_1+\sigma_2^
{\gamma}s_2+\sigma_3^{\gamma}s_3],
\label{bdm}
\end{equation}
where $\sigma_1^{\gamma}, \;\sigma_2^{\gamma}, \;\sigma_3^{\gamma}$  
denote Pauli  matrices, whose rows and columns are labelled by $\hat{\boldsymbol  
u}_{\pm 1}$. We may also note that any arbitrary state of polarization of  
radiation, represented by a point on the Poincare sphere with polar coordinates  
$(\theta_P, \varphi_P)$, may be represented by
\begin{equation}
\hat{\boldsymbol \epsilon}(\alpha, \beta) = \hat{\boldsymbol u}_
{\alpha}\cos{\beta}
+ i\hat{\boldsymbol u}_{\alpha+\pi/2} \sin{\beta}
\end{equation}
with $0 \leq \alpha = \varphi_P /2 < \pi $ and $ -\pi /4 \leq  
\beta=\pi /4 -\theta_P /2
\leq \pi /4$, corresponding to which the differential cross-section is   given by 
\begin{widetext}
\begin{eqnarray}
\frac{d\sigma(\alpha, \beta)}{d\Omega} &=& \frac{q}{4k}\sum_{s_f}[s_f]
^2\sum_{\lambda} \sum_{m_{\lambda}}  
|{\mathcal F}^{\lambda}_{m_{\lambda}}(s_f, 1)
(\cos{\beta} +\sin{\beta})e^{-i\alpha} + 
{\mathcal F}^{\lambda}_{m_{\lambda}}(s_f, -1)
(\cos{\beta} -\sin{\beta})e^{i\alpha}|^2  \\ 
&=&
\frac{q}{4k}\sum_{s_f}[s_f]^2\sum_{\lambda}
\sum_{m_{\lambda}} 
[(1 +\sin{2\beta})|{\mathcal F}^{\lambda} _{m_{\lambda}}
(s_f, 1)|^2 + 
(1 - \sin{2\beta}) |{\mathcal F}^{\lambda}_{m_{\lambda}}(s_f, -1)|^2  
\nonumber\\
&+&
2 \cos{2\beta}\sin{2\alpha} \  \Re [{\mathcal  
F}^{\lambda}_{m_{\lambda}} (s_f, 1)
{\mathcal F}^{\lambda}_{m_{\lambda}}(s_f, -1)^* ]
+2 \cos{2\beta} \sin{2\alpha} 
\ \Im [{\mathcal F}^{\lambda}_{m_{\lambda}} (s_f, 1)
{\mathcal F}^{\lambda}_{m_{\lambda}}(s_f, -1)^* ]],
\end{eqnarray}
\end{widetext}
which readily specializes to give Eqs. \eqref{updc}, \eqref{updcf} and  
\eqref{dcbp}.

\subsection{ General expression for hadron spin observables}
The state of polarization of the target is conveniently specified in  
terms of the initial spin density matrix
\begin{equation}
\rho^i = \frac{1}{2}[1+ {\boldsymbol \sigma}\cdot {\boldsymbol P}]
\label{rhoi},
\end{equation}
where ${\boldsymbol \sigma}$ denote Pauli spin matrices for the  
nucleon and 
${\boldsymbol P}$ its polarization. Using Eqs. \eqref{bdm} and  
\eqref{rhoi}, 
the state of hadron polarization in the final state is, in general,  
described by the 
final spin density matrix
\begin{equation}
\rho^f = \sum_{\mu,\mu^{\prime} = -1,1}{\mathcal F}(\mu)\rho^i 
\rho^{\gamma}_{\mu \mu^{\prime}}{\mathcal F}^{\dagger}(\mu^{\prime}) 
\label{rhof}
\end{equation}
where $\rho^{\gamma}_{\mu \mu^{\prime}}$ denote elements of Eq. 
\eqref{bdm} and ${\mathcal F}^{\dagger}(\mu)$, which denotes the 
hermitian conjugate of Eq. \eqref{ra} with respect to hadron spin  
states may be written as 
\begin{eqnarray}
{\mathcal F}^\dagger (\mu) &=& \sum_{s_f}
\sum_{\lambda} 
(-1)^{{\textstyle{\frac{1}{2}}}-s_f}[s_f][s_i]^{-1}\nonumber \\  
&\times&
\big(S^{\lambda}(s_i,s_f)\cdot {\mathcal F}^{\dagger \lambda}
(s_f, \mu)\big),
\end{eqnarray}
where the notation
\begin{equation}
{\mathcal F}_{m_{\lambda}}^{\dagger \lambda}(s_f, \mu) = (-1)^
{m_{\lambda}}
{\mathcal F}_{-m_{\lambda}}^{\lambda}(s_f, \mu)^*
\end{equation}
is used.  The polarized differential cross-section, in general, is  given by
\begin{equation}
\frac{d\sigma}{d\Omega}= \frac{q}{k}{\rm Tr} \rho^f,
\label{pdcs}
\end{equation}
where Tr denotes the trace or spur. The results of \ref{3a} are simply  
particular  cases of Eq. \eqref{pdcs} with ${\boldsymbol P}=0$. Therefore, all the 
analyzing powers and initial spin correlations may be defined in terms  
of Eq.  \eqref{pdcs}, whereas comprehensive information with regard to spin  
transfers as well as final state polarizations and spin correlations are  
contained in Eq.  \eqref{rhof}. To proceed further and to derive explicit formulae for  
the spin  observables in terms of the irreducible tensor amplitudes Eq.  
\eqref{cf},  we introduce the notations
\begin{equation}
\rho^i =  \frac{1}{2}\sum_{\lambda_i =0}^1 \big( S^{\lambda_i}
({\textstyle{\frac{1}{2}}},{\textstyle{\frac{1}{2}}})
\cdot P^{\lambda_i}
\big),
\end{equation}
where
$
P_0^0 =1 \; ; \; P_0^1 = P_z \; ; \; P_{\pm 1}^1 = \mp  
\frac{1}{\sqrt{2}}
[P_x \pm iP_y]
$
and
\begin{eqnarray}
\rho^{f}(\mu,\mu^{\prime})&=&
{\mathcal F}(\mu)\rho_{i}{\mathcal F}^\dagger (\mu^{\prime})
\nonumber \\
&=& 
\sum_{s_{f}, s_{f}^{\prime}} \sum_{\lambda_{f}=|s_{f}- s_{f}
^{\prime}|}^{ s_{f}+ s_{f}^{\prime}}
\big( S^{\lambda_f}
(s_{f}, s_{f}^{\prime})
\cdot t^{\lambda_f}
\big).
\end{eqnarray}
Using the known \cite{gr1} property
\begin{eqnarray} 
(S^{\lambda_{2}}(s_3,s_2)\otimes  
S^{\lambda_{1}}(s_2,s_1))^{\lambda}_{m_{\lambda}}
 & =& (-1)^{\lambda_{1}+\lambda_{2}-\lambda}
[\lambda_{1}][\lambda_{2}]\nonumber\\
&\times& [s_{2}]W(s_{1}\lambda_{1}s_{3}\lambda_{2};s_{2}\lambda  
)\nonumber \\
&\times&
 S^{\lambda}_{m_{\lambda }} (s_{3},s_{1}),
\label{spin}
\end{eqnarray}
of the irreducible tensor operators and standard Racah algebra, we  
have
\begin{equation} 
\label{tlm}
t^{\lambda_{f}}_{\mu_{f}} =
\sum_{\lambda_{i},\Lambda , \lambda , \lambda^{\prime}}
G\ (
P^{\lambda_{i}} \otimes 
B^{\Lambda}(\lambda s_{f};\lambda^{\prime} s_{f}^{\prime})_
{\mu\mu^{\prime}})^{\lambda_{f}}_{\mu_{f}},
\end{equation}
where $B^{\Lambda}_{m_{\Lambda }}(\lambda s_{f};\lambda^{\prime}
 s_{f}^{\prime})_{\mu\mu^{\prime}}$ 
denote the bilinear irreducible tensors
\begin{equation}
\label{bit} 
B^{\Lambda}_{m_{\Lambda }}(\lambda s_{f};\lambda^{\prime}
s_{f}^{\prime})_{\mu\mu^{\prime}} =
({\mathcal F}^{\lambda}(s_f, \mu) \otimes
{\mathcal F}^{\dagger\,\lambda^{\prime}}(s_f^{\prime}, 
\mu^{\prime}))^{\Lambda}_{m_{\Lambda }}
\end{equation}
of rank $\Lambda$ and the geometrical factors $G$ are given by
\begin{eqnarray} 
\label{g}
G & = &
\frac{1}{\sqrt{2}}\  (-1)^{s_f^{\prime} -{\textstyle{\frac{1}{2}}}}
\ [s_f^{\prime}][\lambda_{i}][\lambda][\lambda^{\prime}][\Lambda]
\nonumber \\
&&
\times\ \sum_{\lambda^{\prime \prime}=|\lambda_i - 
\lambda^\prime |}^{\lambda_i + \lambda^\prime }
(-1)^{\lambda_f - \lambda^\prime - \lambda^{\prime \prime} +
\Lambda} \ [\lambda^{\prime \prime}]^{2} 
\nonumber \\
&&
\times \ 
W(s_{f}^{\prime}\lambda^{\prime}{\textstyle{\frac{1}{2}}}\lambda_{i} 
; {\textstyle{\frac{1}{2}}}\lambda^{\prime\prime } )
W(s_{f}^{\prime}\lambda^{\prime \prime}s_{f}\lambda ; 
{\textstyle{\frac{1}{2}}}\lambda_{f} )
\nonumber \\
&&
\times \  W(\lambda_{i}\lambda^{\prime}\lambda_{f}\lambda ;
\lambda^{\prime\prime }\Lambda ).
\end{eqnarray}
We may thus express $\rho^f$ in terms of its elements
\begin{eqnarray} 
\rho^{f}_{s_{f}m_{f};s_{f}^{\prime}m_{f}^{\prime}} &=&
\sum_{\lambda_{f}}
(-1)^{\mu_{f}}
C(s_{f}^{\prime}\lambda_{f}s_{f};m_{f}^{\prime}-\mu_{f}m_{f})\nonumber  
\\&&
\times \ [\lambda_{f}] T^{\lambda_{f}}_{\mu_{f}},
\label{rf}
\end{eqnarray}
where the $T^{\lambda_{f}}_{\mu_{f}}$ are given in general by
\begin{equation} 
T^{\lambda_{f}}_{\mu_{f}} =\sum_{\mu,\mu^{\prime}=-1,1}
t^{\lambda_{f}}_{\mu_{f}}\rho^{\gamma}_{\mu \mu^{\prime}},
\label{tf}
\end{equation}
in terms of $t^{\lambda_{f}}_{\mu_{f}}$ given by Eq. \eqref{tlm} and 
$\rho^\gamma$ specified by 
Eq. \eqref{bdm}.
If the target is unpolarized, the $T^{\lambda_{f}}_{\mu_{f}}$ reduce  to
\begin{eqnarray} 
T^{\lambda_{f}}_{\mu_{f}} &=&\frac{1}{2}
\sum_{\lambda , \lambda^{\prime}} (-1)^{s_{f}^{\prime}-
{\textstyle{\frac{1}{2}}}}
 [\lambda] [\lambda^\prime] [s_{f}^{\prime}]\nonumber \\
&&
\times \ W(s_{f}^{\prime}\lambda^{\prime}s_{f}\lambda ; 
{\textstyle{\frac{1}{2}}}\lambda_{f} )
B^{\lambda_{f}}_{\mu_{f}}(\lambda s_{f};\lambda^{\prime}  
s_{f}^{\prime}),
\label{bt}
\end{eqnarray}
where $B^{\lambda_{f}}_{\mu_{f}}(\lambda s_{f};\lambda^{\prime}  
s_{f}^{\prime})$
are given in terms of \eqref{bit} through
\begin{equation} 
B^{\Lambda}_{m_{\Lambda}}(\lambda s_{f};\lambda^{\prime} 
s_{f}^{\prime})
=\sum_{\mu,\mu\prime=-1,1}
B^{\Lambda}_{m_{\Lambda }}(\lambda s_{f};\lambda^{\prime} 
s_{f}^{\prime})_{\mu\mu^{\prime}}
\rho^{\gamma}_{\mu \mu\prime},
\label{blm}
\end{equation}
with $\Lambda = \lambda_f$ and $m_{\Lambda} = \mu_f$.

If the beam is also unpolarized, we may replace $\rho^{\gamma}_{\mu  
\mu^{\prime}}$ 
by ${\textstyle{\frac{1}{2}}} \delta_{\mu \mu^{\prime}}$, so that 
we have 
\begin{equation} 
B^{\Lambda}_{m_{\Lambda}}(\lambda s_{f};\lambda^{\prime}  
s_{f}^{\prime})_{0}
={\textstyle{\frac{1}{2}}}
\sum_{\mu=-1,1}
B^{\Lambda}_{m_{\Lambda}}(\lambda s_{f};\lambda^{\prime}  
s_{f}^{\prime})_{\mu\mu}.
\label{bf}
\end{equation}
We may note also that Eq. \eqref{blm} may then be written as
\begin{eqnarray} 
B^{\Lambda}_{m_{\Lambda}}(\lambda s_{f};\lambda^{\prime} 
s_{f}^{\prime})&=&
{\textstyle{\frac{1}{2}}}[
B^{\Lambda}_{m_{\Lambda}}(\lambda s_{f};\lambda^{\prime} 
s_{f}^{\prime})_{0} \nonumber \\
&+&\sum_{i=1}^{3} s_{i}\ 
B^{\Lambda}_{m_{\Lambda}}(\lambda s_{f};\lambda^{\prime}
 s_{f}^{\prime})_{i}],
\label{bl}
\end{eqnarray}
where $s_{1}, s_{2}, s_{3}$ denote the Stokes parameters  
characterizing 
the state of polarization of the beam and
\begin{equation} 
B^{\Lambda}_{m_{\Lambda}}(\lambda s_{f};\lambda^{\prime} 
s_{f}^{\prime})_{i} =
\sum_{\mu,\mu\prime=-1,1}
B^{\Lambda}_{m_{\Lambda }}(\lambda s_{f};\lambda^{\prime}
 s_{f}^{\prime})_{\mu\mu^{\prime}}
(\sigma_{i})_{\mu\mu^{\prime}}.
\label{sp}
\end{equation}

It may be noted that $s_0$ representing the intensity of the beam is  
chosen as 1.
It is worth noting that the bilinears $B^{\Lambda}_{m_{\Lambda  
}}(\lambda
 s_{f};\lambda^{\prime} s_{f}^{\prime})_{\mu\mu^{\prime}}$ 
 are known individually, if the 
$B^{\Lambda}_{m_{\Lambda }}(\lambda s_{f};\lambda^{\prime} 
s_{f}^{\prime})_{i=0,1,2,3}$ are determined empirically.

\subsection{Target analyzing powers}

The differential cross-section Eq. \eqref{pdcs} for a polarized target  
may 
readily be expressed in the form
\begin{equation} 
\frac{d\sigma}{d\Omega} =
\frac{d\sigma_0}{d\Omega}
[1+{\boldsymbol A}\cdot {\boldsymbol P}],
\end{equation}
where the target analyzing power ${\boldsymbol A}$ is given in terms  
of its 
spherical components 
$ A_0^1 = A_z \; ; \; A_{\pm 1}^1 = \mp \frac{1}{\sqrt{2}}
[A_x \pm iA_y]$ by
\begin{eqnarray} 
\frac{d\sigma_0}{d\Omega}
A_{\nu}^{1} &=&\frac{1}{\sqrt{2}}\ 
\frac{q}{k}\sum_{s_{f}}
[s_{f}]^{2}
\sum_{\lambda ,\lambda^{\prime}}
(-1)^{\lambda}
[\lambda]
[\lambda^{\prime}]
\nonumber \\
&&
\times \ W({\textstyle{\frac{1}{2}}} 1 s_{f} \lambda ;
{\textstyle{\frac{1}{2}}}\lambda^{\prime})
B^{1}_{\nu}(\lambda s_{f};\lambda^{\prime} s_{f}).
\end{eqnarray}

\subsection{Polarization in the final state}

Starting with Eq. \eqref{rf} and effecting a change of basis from the  
channel 
spin states to the individual spin states of the meson and the nucleon  
in the 
final state through
\begin{eqnarray} 
\rho^{f}_{m_{s}m_{N}; m_{s}^{\prime}m_{N}^{\prime}}
&=&
\sum_{s_{f},s_{f}^{\prime}}
C(s {\textstyle{\frac{1}{2}}}
 s_{f};m_{s}m_{N}m_{f})
\rho^{f}_{s_{f}m_{f}; s_{f}^{\prime}m_{f}^{\prime}}
\nonumber \\
&&
\times \ C(s {\textstyle{\frac{1}{2}}}
 s_{f}^{\prime};m_{s}^{\prime}m_{N}^{\prime} m_{f}^{\prime}),
\end{eqnarray}
we may express $\rho^f$ in the form
\begin{equation} 
\rho^{f} =\frac{1}{2[s]^{2}}
\sum_{\lambda_{s},\lambda_{N},\lambda_{f}}
( (S^{\lambda_{s}}(s,s) \otimes 
S^{\lambda_{N}}({\textstyle{\frac{1}{2}}}, {\textstyle{\frac{1}{2}}} )
 )^{\lambda_{f}}\cdot E^{\lambda_{f}})
\end{equation}
where
\begin{eqnarray} 
E^{\lambda_{f}}_{\mu_{f}}&=&
\sqrt{2}\,[s]\sum_{s_{f},s_{f}^{\prime}}
(-1)^{s_{f}^{\prime} - s_{f}}
(-1)^{\lambda_{s}+\lambda_{N}-\lambda_{f}}
\nonumber \\
&&
\times \ [s_{f}^{\prime}][s_{f}]^{2} [\lambda_{s}]
[\lambda_{N}]
\left\{ \begin{matrix}
s & {\textstyle{\frac{1}{2}}}
& s_{f}^{\prime}\\[.2cm]
s & {\textstyle{\frac{1}{2}}}
& s_{f}\\[.2cm]
\lambda_{s}&   \lambda_{N}& \lambda_{f}
\end{matrix} \right \}
T^{\lambda_{f}}_{\mu_{f}},
\label{en}
\end{eqnarray}
and $\{\}$ denotes the Wigner 9j symbol \cite{rusi}. 
The irreducible tensors $E^{\lambda_{f}}_{\mu_{f}}$ 
describe clearly that the meson and nucleon spins are entangled in the 
final state through Eq. \eqref{en}. It is clear from Eqs.  \eqref{bt}  
to 
\eqref{bf} that the entanglement persists even if the target and beam  
are 
unpolarized. If no observations are made on the spin state of the  
recoil nucleon in 
the final state, the state of polarization of the emitted meson is  
characterized by 
the density matrix $\rho^{s}$, whose elements are given by
\begin{equation} 
\rho^{s}_{m_{s}m_{s}^{\prime}} =
\sum_{m_{N}=-{\textstyle{\frac{1}{2}}}}
^{{\textstyle{\frac{1}{2}}}}
\rho^{f}_{m_{s}m_{N}; m_{s}^{\prime}m_{N}} .
\label{rhos}
\end{equation}

Likewise, if no observations are made on the meson spin state, the  
recoil nucleon polarization is specified by the density matrix $\rho^{r}$ 
whose elements are given by
\begin{equation} 
\rho^{r}_{m_{N}m_{N}^{\prime}} =
\sum_{m_{s}=-s}^{s}
\rho^{f}_{m_{s}m_{N}; m_{s}m_{N}^{\prime}} .
\label{rr}
\end{equation}

\subsection{Recoil nucleon polarization and nucleon spin transfer}

Expressing $\rho^{r}$ given by Eq. \eqref{rr} in the form
\begin{equation} 
\rho^r = \frac{{\rm Tr} \rho^r}{2}
[1+ {\boldsymbol \sigma}\cdot {\boldsymbol R}],
\end{equation}
the recoil nucleon polarization ${\boldsymbol R}$ is given in terms of  
its spherical components 
$ R_0^1 = R_z \; ; \; R_{\pm 1}^1 = \mp \frac{1}{\sqrt{2}}
[R_x \pm i R_y]$ by
\begin{equation} 
\frac{d\sigma_0}{d\Omega}R^{1}_{\mu_{f}}=
\frac{q}{k}
\sum_{s_{f},s_{f}^{\prime}}
G_{r}
T^{1}_{\mu_{f}}
\label{rp}
\end{equation}
where the geometrical factor $G_{r}$ is given by 
\begin{equation} 
G_{r} = \sqrt{2}[s_{f}^{\prime}][ s_{f}]^{2}
W(s{\textstyle{\frac{1}{2}}}  
s_{f}1;s_{f}^{\prime}{\textstyle{\frac{1}{2}}}).
\label{gr}
\end{equation}

If the target is polarized, it is clear from Eqs. \eqref{tlm}  and  
\eqref{tf} 
that $R^{1}_{\mu_{f}}$ is dependent on ${\boldsymbol P}$ and this 
can be brought out by expressing Eq. \eqref{rp} in the form
\begin{equation}
\frac{d\sigma_0}{d\Omega}R^{1}_{\mu_{f}}= 
\sum_{\mu_{i}}
{\cal R}^{\prime\prime}_{\mu_{f}\mu_{i}}
P^{1}_{\mu_{i}}
 \end{equation}
in terms of the nucleon spin transfers ${\cal  
R}^{\prime\prime}_{\mu_{f}\mu_{i}}$ 
which are readily given by
\begin{eqnarray} 
{\cal R}^{\prime\prime}_{\mu_{f}\mu_{i}}&=&
\frac{q}{k}
\sum_{\Lambda} C(1\Lambda 1;\mu_{i}m_{\Lambda}\mu_{f})
\sum_{s_{f},s_{f}^{\prime}} G_{r}\nonumber\\
&&
\times \ \sum_{\lambda,\lambda^{\prime}} G\ 
B^{\Lambda}_{m_{\Lambda }}(\lambda s_{f};\lambda^{\prime}  
s_{f}^{\prime}),
\end{eqnarray}
where $G_{r}, G$ and $B^{\Lambda}_{m_{\Lambda }}(\lambda 
s_{f};\lambda^{\prime} s_{f}^{\prime})$
are given respectively by Eqs. \eqref{gr}, \eqref{g} and \eqref{blm}.
It may be noted that the recoil nucleon is polarized even in the  
absence of target polarization.

\subsection{Meson polarization and target nucleon to meson spin  
transfer}

Using $\tau^{\lambda_{s}}_{\mu_{s}} =  
S^{\lambda_{s}}_{\mu_{s}}(s,s)$, 
we may express $\rho^s$ 
given by Eq. \eqref{rhos} in the form
\begin{equation} 
\rho^s = \frac{{\rm Tr} \rho^s}{2s+1}
\sum_{\lambda_{s}=0}^{2s}(\tau^{\lambda_{s}}\cdot
t^{\lambda_{s}}) ,
\end{equation}
in terms of Fano statistical tensors $t^{\lambda_{s}}_{\mu_{s}}$
of rank $\lambda_{s}$ which are given by
\begin{equation} 
\frac{d\sigma_0}{d\Omega}
t^{\lambda_{s}}_{\mu_{s}}=\frac{q}{k}
\sum_{s_{f},s_{f}^{\prime}}
G_{s} T^{\lambda_{s}}_{\mu_{s}}, 
\label{fst}
\end{equation}
where the geometrical factor $G_s$ is given by
\begin{equation} 
G_{s} = (-1)^{s_{f}-s_{f}^{\prime}} [s_{f}]^{2}[s_{f}^{\prime}]
W({\textstyle{\frac{1}{2}}}ss_{f}\lambda_{f}; s_{f}^{\prime}s).
\end{equation}
Noting from Eqs. \eqref{tlm} and \eqref{tf}, that  
$T^{\lambda_{s}}_{\mu_{s}}$ 
are dependent on ${\boldsymbol P}$, when the target is polarized, we  
may define the target 
nucleon to meson spin transfer 
${\mathcal T}^{\lambda_{s} 1}_{\mu_{s} \mu_{i}}$ through
\begin{equation} 
\frac{d\sigma_0}{d\Omega}
t^{\lambda_{s}}_{\mu_{s}}=
\sum_{\mu_{i}}
{\mathcal T}^{\lambda_{s} 1}_{\mu_{s} \mu_{i}}
P^{1}_{\mu_{i}}
\end{equation}
so that
\begin{eqnarray} 
{\mathcal T}^{\lambda_{s} 1}_{\mu_{s} \mu_{i}} &=&
\frac{q}{k}
\sum_{\Lambda}
C(1\Lambda\lambda_{s};\mu_{i}m_{\Lambda}\mu_{s})\nonumber\\
&&
\times \ \sum_{s_{f},s_{f}^{\prime}}
G_{s}
\sum_{\lambda,\lambda^{\prime}}
G\  B^{\Lambda}_{m_{\Lambda }}(\lambda s_{f};\lambda s_{f}^{\prime}).
\end{eqnarray}

The meson is polarized even if the target is unpolarized. In such a  
case, the  Fano statistical tensors $t^{\lambda_{s}}_{\mu_{s}}$ 
characterizing the meson polarization are given by Eq. \eqref{fst}, 
where use is made of Eq. \eqref{bt} for 
$T^{\lambda_{s}}_{\mu_{s}}$  on the right hand side of Eq.  
\eqref{fst}.

\section{Summary and Outlook}

Photo production of mesons with arbitrary spin-parity $s^{\pi}$ is  
described  in terms of a set of $4(2s+1)$ independent irreducible tensor  
amplitudes, which are expressible in terms of a single compact formula given by Eq.  
\eqref{cf}  in terms of the `magnetic' and `electric' multipole amplitudes 
$ {\mathcal M}^j_{ls_f;L} $ and  $ {\mathcal E}^j_{ls_f;L} $  
respectively introduced through Eq. \eqref{pwmpa}. All  hadron spin observables  
viz., the target analyzing power,  meson and recoil nucleon polarizations, target to 
meson and target to   recoil nucleon  spin transfers and spin correlations 
and beam analyzing powers  have been  expressed in terms of the bilinear tensors 
$B^{\Lambda}_{m_{\Lambda }}(\lambda s_{f};\lambda^{\prime} 
 s_{f}^{\prime} )_{\mu \mu'}$ of rank $\Lambda = 0, ..., 2(s+1)$
defined by Eq. \eqref{bit}.  The unpolarized differential cross-section itself
 is clearly proportional to 
$B^{0}_{0}(\lambda s_{f};\lambda s_{f})_0$, while hadron
spin observables in experiments with an unpolarized beam are  given by 
$B^{\Lambda}_{m_{\Lambda }}(\lambda s_{f};\lambda^{\prime}  
s_{f}^{\prime} )_{0}$  in accordance with Eq. \eqref{bf}. 
It is also clear from Eqs. \eqref{bl} and \eqref{sp} that each element 
$B^{\Lambda}_{m_{\Lambda }}(\lambda s_{f};\lambda^{\prime}  
s_{f}^{\prime} )_{\mu\mu^{\prime} }$ for
$\mu, \mu^{\prime}  = \pm 1$ is known if the four 
$B^{\Lambda}_{m_{\Lambda }}(\lambda s_{f};\lambda^{\prime}  
s_{f}^{\prime})_{i}$ for
$i = 0,1,2,3$  with unpolarized and appropriate partially polarized beams 
are measured for a given $\Lambda$ and $m_{\Lambda}$.  
Thus our approach employing  irreducible tensor amplitudes  leads to a  
systematic  and elegant procedure to analyze experimental data on all spin and  
polarization  observables including beam analyzing powers and the 
differential cross-section for photo production of mesons with  
arbitrary  spin-parity $s^\pi$.  Further  work is in progress. 

\begin{acknowledgments}
One of us (G.R.) is grateful to Professors B.V. Sreekantan, R. Cowsik,  
J.H. Sastry,  R. Srinivasan and S.S. Hasan for facilities provided for research at  
the  Indian Institute of Astrophysics and another (J.B.) acknowledges much 
encouragement for research given by the Principal Dr. T.G.S. Moorthy  
and  the Management of K.S. Institute of Technology.
\end{acknowledgments}

\appendix
\section {Connection between irreducible tensor amplitudes and CGLN  
amplitudes for pion production}
 
For the particular case of pseudoscalar meson photo production, the  
final state channel spin can take only one value namely
$s_{f}={\textstyle\frac{1}{2}}$ since $s=0$. It is clear that 
${\mathcal F}^{\lambda}_{m_{\lambda}}
({\textstyle\frac{1}{2}}, \mu)$ for $ \lambda = 0,1$ represent 
respectively the nucleon spin independent and spin dependent 
amplitudes. The CGLN amplitudes ${\mathcal F}_i,\; i=1, ...,4$ in  
terms of which the reaction amplitude ${\mathcal F}$ is expressed by Eq. (7.2) of   
\cite{che} are well known. Re-writing Eq. (7.2) of \cite{che} conveniently as 
\begin{equation}
{\mathcal F} = L + i\;{\boldsymbol \sigma} \cdot {\boldsymbol K},
\end{equation}
in terms of the nucleon spin independent and spin dependent amplitudes
$L$ and ${\boldsymbol K}$ respectively, we may note 
\begin{eqnarray}
\label{spinin}
L& =& \hat{\boldsymbol q}\cdot (\hat{\boldsymbol k} \times \hat 
{\boldsymbol \epsilon})\;{\mathcal F}_2  \\
\label{spinde}
{\boldsymbol K}&=& \hat{\boldsymbol \epsilon}\;({\mathcal F}_1 
-{\mathcal F}_2 \cos{\theta} )  \nonumber \\
&& + (\hat{\boldsymbol q} \cdot 
\hat{\boldsymbol \epsilon})[\hat{\boldsymbol q}\;{\mathcal F}_4 
+ \hat{\boldsymbol k}\;({\mathcal F}_2+{\mathcal F}_3)],
\end{eqnarray}
where $\hat{\boldsymbol \epsilon}$ denotes photon polarization which
is orthogonal to $\hat{\boldsymbol k}$.  We may define $L(\mu)$ 
and $K^1_{m_{\lambda}}(\mu)$ by substituting 
$\hat{\boldsymbol \epsilon} = \hat{\boldsymbol u}_{\mu}$
in (\ref{spinin}) and (\ref{spinde}) respectively after expressing  
${\boldsymbol K}$
in terms of its spherical components $K^1_0 = K_z \; ; K^1_{\pm 1}=  
\mp \frac{1}
{\sqrt{2}}(K_x\pm iK_y))$. We thus have 
\begin{eqnarray}
\label{x1}
{\mathcal F}_{0}^{0}({\textstyle{\frac{1}{2}}},\pm 1)&=& L(\pm 1) =  
\mp  \frac{i}{\sqrt{2}}{\mathcal F}_2 \sin{\theta}\\
{\mathcal F}_{0}^{1}({\textstyle{\frac{1}{2}}},\pm 1)&=&  
i\;K^1_0(\pm1)  \nonumber \\ &=&
\frac{i}{\sqrt{2}}\sin{\theta} [{\mathcal F}_2 
+{\mathcal F}_3 + {\mathcal F}_4\cos{\theta}] \\
{\mathcal F}_{-1}^{1}({\textstyle{\frac{1}{2}}},\pm1 )&=&  
i\;K^1_{-1}(\pm1) 
\nonumber \\ &=&  \pm i\;[  {\mathcal F}_{1} - {\mathcal  
F}_{2}\cos{\theta} + 
{\mathcal F}_{4} \frac{\sin^{2}{\theta}}{2}] \\
{\mathcal F}_{1}^{1}({\textstyle{\frac{1}{2}}},
\pm 1)&=& i\;K^1_1(\pm1) = \mp
i {\mathcal F}_4\frac{ \sin^{2}{\theta}}{2},
\label{lst}
\end{eqnarray} 
where the irreducible tensor amplitudes are explicitly 
given,  on using Eq. (\ref{cf}), with $\varphi =0$ in Madison Frame by 
\begin{widetext}
\begin{eqnarray}
\label{a1}
{\mathcal F}_{0}^{0}({\textstyle\frac{1}{2}},\pm 1)&=&  
\mp \frac{4\pi}{\sqrt{2}}\sin{\theta}
\sum_{l} P_{l}^{\ \prime}\frac{1}{\sqrt{l(l+1)}}[(l+1)
{\mathcal M}^{l+{\textstyle\frac{1}{2}}}_{l,l}+
l{\mathcal M}^{l-{\textstyle\frac{1}{2}}}_{l,l}] \\
{\mathcal F}_{0}^{1}({\textstyle\frac{1}{2}},\pm 1)&=& 
\frac{4\pi}{\sqrt{2}}\sin{\theta}\sum_{l}
P_{l}^{\ \prime}[-\frac{{\mathcal  
M}^{l+{\textstyle\frac{1}{2}}}_{l,l}}{\sqrt{l(l+1)}}
+\frac{{\mathcal M}^{l-{\textstyle\frac{1}{2}}}_{l,l}}{\sqrt{l(l+1)}}
-\sqrt{\frac{(l+2)}{(l+1)}} {\mathcal  
E}^{l+{\textstyle\frac{1}{2}}}_{l,l+1}
+\sqrt{\frac{(l-1)}{l}} {\mathcal  
E}^{l-{\textstyle\frac{1}{2}}}_{l,l-1}]
\\
{\mathcal F}_{-1}^{1}({\textstyle\frac{1}{2}},\pm 1)&=& \pm  
2\pi\sum_{l} P_{l}
[\sqrt{l(l+1)}( {\mathcal M}^{l+{\textstyle\frac{1}{2}}}_{l,l}-
{\mathcal M}^{l-{\textstyle\frac{1}{2}}}_{l,l})
-\sqrt{(l+2)(l+1)} {\mathcal E}^{l+{\textstyle\frac{1}{2}}}_{l,l+1}
-\sqrt{l(l-1)} {\mathcal E}^{l-{\textstyle\frac{1}{2}}}_{l,l-1}]
\\ \label{a4}
{\mathcal F}_{1}^{1}({\textstyle\frac{1}{2}},\pm 1)&=& \pm  
2\pi\sin^{2}{\theta}
\sum_{l} P_{l}^{\ \prime\prime} 
[-\frac{{\mathcal M}^{l+{\textstyle\frac{1}{2}}}_{l,l}}{\sqrt{l(l+1)}}
+\frac{{\mathcal M}^{l-{\textstyle\frac{1}{2}}}_{l,l}}{\sqrt{l(l+1)}}
-\frac{{\mathcal  
E}^{l+{\textstyle\frac{1}{2}}}_{l,l+1}}{\sqrt{(l+2)(l+1)}}
-\frac{{\mathcal  
E}^{l-{\textstyle\frac{1}{2}}}_{l,l-1}}{\sqrt{l(l-1)}}].
\label{x4}
\end{eqnarray}
\end{widetext}
Comparing \eqref{x1} to \eqref{lst} with \eqref{a1} to 
\eqref{a4} and using Eqs. (7.3) to (7.6) of CGLN  \cite{che}, for 
${\mathcal F}_i,\; i=1, ...,4$, we may identify 
\begin{eqnarray}
{\mathcal M}^{l\pm{\textstyle\frac{1}{2}}}_{l,L}&=&  
i\frac{\sqrt{L(L+1)}}{4\pi}M_{l\pm}\\
{\mathcal E}^{l\pm{\textstyle\frac{1}{2}}}_{l,L}  
&=&-i\frac{\sqrt{L(L+1)}}{4\pi}E_{l\pm}
\end{eqnarray}
 in terms of the multipole amplitudes, $M_{l\pm}$ and $E_{l\pm}$ of
CGLN \cite{che}.

The photo production amplitudes for specific reactions are readily  obtained using 
Eq. (\ref{ispa}) 
\begin{eqnarray}
\label{a14}
 \langle n \pi^+|{\mathcal F}|\gamma p \rangle &=&  
-{\textstyle\frac{\sqrt{2}}{3}} 
 [{\mathcal F}^{1\textstyle{\frac{1}{2}}}-{\mathcal  
F}^{1\textstyle{\frac{3}{2}}}
  + \sqrt{3}{\mathcal F}^{0\textstyle{\frac{1}{2}}}]\\
  \langle p \pi^0 |{\mathcal F}|\gamma p \rangle &=&  
{\textstyle\frac{1}{3}}  
 [2{\mathcal F}^{1\textstyle{\frac{3}{2}}}+{\mathcal  
F}^{1\textstyle{\frac{1}{2}}}
  + \sqrt{3}{\mathcal F}^{0\textstyle{\frac{1}{2}}}]\\
  \langle p \pi^-|{\mathcal F}|\gamma n \rangle &=&  
{\textstyle\frac{\sqrt{2}}{3}} 
 [{\mathcal F}^{1\textstyle{\frac{3}{2}}}-{\mathcal  
F}^{1\textstyle{\frac{1}{2}}}
  + \sqrt{3}{\mathcal F}^{0\textstyle{\frac{1}{2}}}]\\
  \label{a17}
  \langle n \pi^0 |{\mathcal F}|\gamma n \rangle &=&  
{\textstyle\frac{1}{3}}  
 [2{\mathcal F}^{1\textstyle{\frac{3}{2}}}+{\mathcal  
F}^{1\textstyle{\frac{1}{2}}}
  - \sqrt{3}{\mathcal F}^{0\textstyle{\frac{1}{2}}}].
\end{eqnarray}
It may be noted that if the different charge states of the pion  are  
defined in the real three  dimensional isospin space through
\begin{equation}
\pi^0 = \pi_z \; ; \; \pi^{\pm} = \frac{1}{\sqrt{2}}(\pi_x \pm  
i\;\pi_y),
\end{equation}
an extra overall minus sign has to be attached to (\ref{a14}), since the three charged
state of the pion are identified  through $|1 \nu_m \rangle , \nu_m = 0, \pm 1$ 
in deriving (\ref{a14}) to (\ref{a17}). We may then compare the above with 
\begin{eqnarray}
\langle n \pi^+|{\mathcal F}|\gamma p \rangle &=& \sqrt{2}[{\mathcal  
F}^{(-)}+ {\mathcal F}^{(0)}]\\ 
  \langle p \pi^0 |{\mathcal F}|\gamma p \rangle &=& {\mathcal  
F}^{(+)}+ {\mathcal F}^{(0)}\\
  \langle p \pi^-|{\mathcal F}|\gamma n \rangle &=&  
-\sqrt{2}[{\mathcal F}^{(-)}- {\mathcal F}^{(0)}]\\
  \langle n \pi^0 |{\mathcal F}|\gamma n \rangle &=& {\mathcal  
F}^{(+)}-{\mathcal F}^{(0)}
\end{eqnarray}
in terms of  the traditional isospin amplitudes of CGLN and hence  
identify 
\begin{eqnarray}
  {\mathcal F}^{(0)} &=&\frac{1}{\sqrt{3}}{\mathcal  
F}^{0\textstyle{\frac{1}{2}}}\\
 {\mathcal F}^{(+)} &=& \textstyle{\frac{1}{3}}[{\mathcal  
F}^{1\textstyle{\frac{1}{2}}} + 
 2{\mathcal F}^{1\textstyle{\frac{3}{2}}}]\\
{\mathcal F}^{(-)} &=& \textstyle{\frac{1}{3}}[{\mathcal  
F}^{1\textstyle{\frac{1}{2}}} - 
 {\mathcal F}^{1\textstyle{\frac{3}{2}}}].
\end{eqnarray}

\section{Helicity amplitudes}

The helicity amplitudes may be defined through 
\begin{equation}
H_{\mu_s \mu_f ; \mu_i \mu} \equiv \langle s \mu_s ;  
{\textstyle{\frac{1}{2}}} \mu_f|
{\mathcal F}|{\textstyle{\frac{1}{2}}} \mu_i ; 1 \mu \rangle
\end{equation}
where $\mu$ and $\mu_s$ denote the photon and meson spin projections  
along 
${\boldsymbol k}$ and ${\boldsymbol q}$ respectively, while 
$|{\textstyle{\frac{1}{2}}}\mu_{i}\rangle$ and 
$|{\textstyle{\frac{1}{2}}}\mu_{f}\rangle$
denote  initial and final helicity states with conventional phases  
following 
\cite{jac} and having spin projections $\mu_{i}$ and $\mu_{f}$ along 
$-{\boldsymbol k}$ and $-{\boldsymbol q}$ respectively.
We may note that the helicity eigenstate of the photon is given by 
\begin{equation}
|1\mu\rangle \equiv \hat{\boldsymbol \xi}_{\mu}=
-\mu\hat{\boldsymbol u}_{\mu};\ \mu=\pm 1,
\end{equation}
following Eq. \eqref{uv}. We also observe that 
\begin{equation}
|s \mu_s ; {\textstyle{\frac{1}{2}}} \mu_f \rangle = \sum_{s_f} C(s 
{\textstyle{\frac{1}{2}}} s_f ; \mu_s -\mu_f \mu_f') |(s  
{\textstyle{\frac{1}{2}}})
s_f \mu_f' \rangle
\end{equation}
in terms of channel spin states $|(s {\textstyle{\frac{1}{2}}}) s_f  
\mu_f' \rangle$
which are expressed as 
\begin{equation}
|(s {\textstyle{\frac{1}{2}}}) s_f \mu_f' \rangle = \sum_{m_f}  
d^{s_f}_{m_f \mu_f'}
(\theta)|(s {\textstyle{\frac{1}{2}}}) s_f m_f \rangle
\end{equation}
with respect to ${\boldsymbol q}$ as the quantization axis. 
The polar co-ordinates of ${\boldsymbol q}$ are
$(q,\theta,0)$ in the Madison frame \cite{mad} 
where the plane containing ${\boldsymbol q}$ and
${\boldsymbol k}$ is chosen as the reaction plane with the
 $z$ axis along ${\boldsymbol k}$.
Thus, we have
\begin{eqnarray}
H_{\mu_s \mu_f ; \mu_i \mu} &=& -\mu \sum_{s_f m_f} 
C(s{\textstyle{\frac{1}{2}}} s_f ; \mu_s -\mu_f \mu_f') d^{\,s_f}_{m_f  
\mu_f'}(\theta) 
\nonumber \\
&& \times \ 
 \sum_{m_i} \langle (s {\textstyle{\frac{1}{2}}}) s_f m_f |
{\mathcal F}(\mu)| {\textstyle{\frac{1}{2}}} m_i \rangle \delta_{m_i,  
-\mu_i}
\label{ha}
\end{eqnarray}
where the matrix elements are expressible as 
\begin{eqnarray}
\langle (s {\textstyle{\frac{1}{2}}}) s_f m_f |
{\mathcal F}(\mu)| {\textstyle{\frac{1}{2}}} m_i \rangle &=&  
\sum_{\lambda}
C({\textstyle{\frac{1}{2}}} \lambda s_f ; m_i -m_{\lambda} m_f)
\nonumber \\
&& \times (-1)^{m_{\lambda}}
[\lambda]
{\mathcal F}^{\lambda}_{m_{\lambda}}(s_f, \mu)
\label{me}
\end{eqnarray}
in terms of the irreducible tensor amplitudes given by Eq. \eqref{cf}  
with $\phi=0$. 
Using \eqref{ha} and Eq. \eqref{sr}, we obtain the symmetry relation,
\begin{equation}
H_{-\mu_s -\mu_f ; -\mu_i -\mu}=-\pi (-1)^{s+\mu_{s}+\mu_{i}-\mu_{f}}
H_{\mu_s \mu_f ; \mu_i \mu}\, .
\end{equation}
In the particular case of psuedoscalar meson photo production 
with $s=\mu_{s}=0, s_{f}={\textstyle{\frac{1}{2}}}$
and $\pi=-1$, the helicity amplitudes are explicitly given by
\begin{eqnarray}
\label{w1}
H_{-{\textstyle\frac{1}{2}}; -{\textstyle\frac{1}{2}}1}
 &=& \sqrt{2}\sin{{\textstyle\frac{\theta}{2}}}
 {\cal F}_{1}^{1}({\textstyle\frac{1}{2}},1) 
\nonumber \\ &-&
\cos{{\textstyle\frac{\theta}{2}}}[{\cal F}_{0}^{0}
({\textstyle\frac{1}{2}},1) + {\cal  
F}_{0}^{1}({\textstyle\frac{1}{2}},1) ]
 \\ 
H_{-{\textstyle\frac{1}{2}}; {\textstyle\frac{1}{2}}1}
 &=& -\sqrt{2}\cos{{\textstyle\frac{\theta}{2}}}
 {\cal F}_{-1}^{1}({\textstyle\frac{1}{2}},1) 
\nonumber \\ &-& 
\sin{{\textstyle\frac{\theta}{2}}}[{\cal F}_{0}^{0}
({\textstyle\frac{1}{2}},1) - {\cal  
F}_{0}^{1}({\textstyle\frac{1}{2}},1) ] 
 \\ 
H_{{\textstyle\frac{1}{2}}; -{\textstyle\frac{1}{2}}1}
 &=& \sqrt{2}\cos{{\textstyle\frac{\theta}{2}}}
 {\cal F}_{1}^{1}({\textstyle\frac{1}{2}},1) 
\nonumber \\ &+&
\sin{{\textstyle\frac{\theta}{2}}}[{\cal F}_{0}^{0}
({\textstyle\frac{1}{2}},1) + {\cal  
F}_{0}^{1}({\textstyle\frac{1}{2}},1)] 
\\
H_{{\textstyle\frac{1}{2}}; {\textstyle\frac{1}{2}}1}
 &=& \sqrt{2}\sin{{\textstyle\frac{\theta}{2}}}
 {\cal F}_{-1}^{1}({\textstyle\frac{1}{2}},1) 
\nonumber \\ &-&
\cos{{\textstyle\frac{\theta}{2}}}[{\cal F}_{0}^{0}
({\textstyle\frac{1}{2}},1) - {\cal  
F}_{0}^{1}({\textstyle\frac{1}{2}},1) ]
\label{w4},
\end{eqnarray}
satisfying 
\begin{equation}
H_{-\mu_f ; -\mu_i -\mu}=(-1)^{\mu_{i}-\mu_{f}}
H_{\mu_f ; \mu_i \mu},
\end{equation}
in exact agreement with Walker \cite{walker}, where the notation 
$H_1, H_2, H_3$ and $H_4$ is used respectively for the 
4 independent helicity amplitudes \eqref{w1},...,\eqref{w4}. 
Using \eqref{a1} to \eqref{x4} in \eqref{me}, 
we readily obtain the helicity amplitudes in terms of the CGLN  
amplitudes as
\begin{eqnarray}
H_1 &=&  
-\frac{1}{\sqrt{2}}\sin{\theta}\cos{{\textstyle\frac{\theta}{2}}}
[{\cal F}_3+{\cal F}_4] \\
H_2 &=&  \sqrt{2} \cos{{\textstyle\frac{\theta}{2}}}[({\cal F}_2-{\cal  
F}_1)\nonumber \\
&+&
\frac{1}{2}(1-\cos{\theta})({{\cal F}_3-{\cal F}_4})]\\
H_3 &=&  
\frac{1}{\sqrt{2}}\sin{\theta}\sin{{\textstyle\frac{\theta}{2}}}
({{\cal F}_3-{\cal F}_4})\\
H_4 &=&  \sqrt{2} \sin{{\textstyle\frac{\theta}{2}}}[({\cal F}_1+{\cal  
F}_2)\nonumber \\
&+&
\frac{1}{2}(1+\cos{\theta})({{\cal F}_3+{\cal F}_4})]
\end{eqnarray}
which are in exact agreement with Walker \cite{walker}.

In the case of vector meson photo production with $s=1$, 
we may once again use \eqref{me} in \eqref{ha} to obtain explicitly 
the 12 independent helicity amplitudes in terms of the irreducible  
tensor amplitudes as
{\small 
\begin{widetext}
\begin{eqnarray}
H_{-1{\textstyle\frac{1}{2}}; -{\textstyle\frac{1}{2}}1} 
&=&\cos{{\textstyle\frac{\theta}{2}}}
[
-\sqrt{{\textstyle\frac{3}{2}}}(1-\cos{\theta})
[{\cal F}_{0}^{1}({\textstyle\frac{3}{2}},1)+{\cal F}_{0}^{2}
({\textstyle\frac{3}{2}},1)] 
- (1+\cos{\theta})
{\cal F}_{2}^{2}({\textstyle\frac{3}{2}},1) ]
\nonumber\\ 
&+&  \sin{{\textstyle\frac{\theta}{2}}}
[-{{\textstyle\frac{\sqrt{3}}{2}}}(1-\cos{\theta})
 [{\cal F}_{-1}^{1}({\textstyle\frac{3}{2}},1)+
 {\textstyle\frac{1}{\sqrt{3}}}{\cal  
F}_{-1}^{2}({\textstyle\frac{3}{2}},1)] 
- {{\textstyle\frac{\sqrt{3}}{2}}}(1+\cos{\theta})
[{\cal F}_{1}^{1}({\textstyle\frac{3}{2}},1)+\sqrt{3}
{\cal F}_{1}^{2}({\textstyle\frac{3}{2}},1)] 
]\\
H_{0{\textstyle\frac{1}{2}}; -{\textstyle\frac{1}{2}}1} 
&=&\cos{{\textstyle\frac{\theta}{2}}}
[ 
\sqrt{{\textstyle\frac{2}{3}}}{\cal  
F}_{1}^{1}({\textstyle\frac{1}{2}},1)
+ \sqrt{{\textstyle\frac{3}{2}}}(1-\cos{\theta})
[{\cal F}_{-1}^{1}({\textstyle\frac{3}{2}},1)+
{\textstyle\frac{1}{\sqrt{3}}}{\cal  
F}_{-1}^{2}({\textstyle\frac{3}{2}},1)] 
+ {\textstyle\frac{1}{\sqrt{6}}}(3\cos{\theta}-1)
\nonumber\\ & \times & [{\cal F}_{1}^{1}({\textstyle\frac{3}{2}},1)+
\sqrt{3}{\cal F}_{1}^{2}({\textstyle\frac{3}{2}},1)]]
\nonumber\\
&+&  \sin{{\textstyle\frac{\theta}{2}}}[  
{\textstyle\frac{1}{\sqrt{3}}}
[{\cal F}_{0}^{0}({\textstyle\frac{1}{2}},1)+{\cal  
F}_{0}^{1}({\textstyle\frac{1}{2}},1)]
+ {\textstyle\frac{1}{\sqrt{3}}}(3\cos{\theta}+1)
[{\cal F}_{0}^{1}({\textstyle\frac{3}{2}},1)+{\cal  
F}_{0}^{2}({\textstyle\frac{3}{2}},1)]
- \sqrt{2}(1+\cos{\theta}) {\cal F}_{2}^{2}({\textstyle\frac{3}{2}},1) 
] \\
H_{1{\textstyle\frac{1}{2}}; -{\textstyle\frac{1}{2}}1} 
&=&\cos{{\textstyle\frac{\theta}{2}}}
[ -\sqrt{{\textstyle\frac{2}{3}}}
[{\cal F}_{0}^{0}({\textstyle\frac{1}{2}},1)+{\cal  
F}_{0}^{1}({\textstyle\frac{1}{2}},1)]
- {\textstyle\frac{1}{\sqrt{6}}}(3\cos{\theta}-1)
[{\cal F}_{0}^{1}({\textstyle\frac{3}{2}},1)+{\cal  
F}_{0}^{2}({\textstyle\frac{3}{2}},1)]
- (1-\cos{\theta}) {\cal F}_{2}^{2}({\textstyle\frac{3}{2}},1) 
]\nonumber\\
&+&  \sin{{\textstyle\frac{\theta}{2}}}
[ {\textstyle\frac{2}{\sqrt{3}}}{\cal  
F}_{1}^{1}({\textstyle\frac{1}{2}},1)
- {\textstyle\frac{\sqrt{3}}{2}}(1+\cos{\theta})
[{\cal F}_{-1}^{1}({\textstyle\frac{3}{2}},1)+
{\textstyle\frac{1}{\sqrt{3}}}{\cal  
F}_{-1}^{2}({\textstyle\frac{3}{2}},1)] 
+ {\textstyle\frac{1}{2\sqrt{3}}}(3\cos{\theta}+1)\nonumber\\ & \times  
&
[{\cal F}_{1}^{1}({\textstyle\frac{3}{2}},1)+
\sqrt{3}{\cal F}_{1}^{2}({\textstyle\frac{3}{2}},1)]
]
\end{eqnarray}
\begin{eqnarray}
H_{-1{\textstyle\frac{1}{2}}; {\textstyle\frac{1}{2}}1} 
&=&\cos{{\textstyle\frac{\theta}{2}}}
[ {\textstyle\frac{\sqrt{3}}{2}}(1-\cos{\theta})
[{\cal F}_{-1}^{1}({\textstyle\frac{3}{2}},1)-\sqrt{3}
{\cal F}_{-1}^{2}({\textstyle\frac{3}{2}},1)]
+{\textstyle\frac{\sqrt{3}}{2}}(1+\cos{\theta})
[{\cal F}_{1}^{1}({\textstyle\frac{3}{2}},1)
-{\textstyle\frac{1}{\sqrt{3}}}{\cal  
F}_{1}^{2}({\textstyle\frac{3}{2}},1)]
]\nonumber\\
&+&  \sin{{\textstyle\frac{\theta}{2}}}
[ \sqrt{{\textstyle\frac{3}{2}}}(1+\cos{\theta})
[{\cal F}_{0}^{1}({\textstyle\frac{3}{2}},1)
-{\cal F}_{0}^{2}({\textstyle\frac{3}{2}},1)]
-(1-\cos{\theta}) {\cal F}_{-2}^{2}({\textstyle\frac{3}{2}},1)
]
 \\
H_{0{\textstyle\frac{1}{2}}; {\textstyle\frac{1}{2}}1} 
&=&\cos{{\textstyle\frac{\theta}{2}}}
[ -{\textstyle\frac{1}{\sqrt{3}}}
[{\cal F}_{0}^{0}({\textstyle\frac{1}{2}},1)
-{\cal F}_{0}^{1}({\textstyle\frac{1}{2}},1)]
- {\textstyle\frac{1}{\sqrt{3}}}(3\cos{\theta}-1)
[{\cal F}_{0}^{1}({\textstyle\frac{3}{2}},1)-{\cal  
F}_{0}^{2}({\textstyle\frac{3}{2}},1)]
+\sqrt{2} (1-\cos{\theta}) {\cal  
F}_{-2}^{2}({\textstyle\frac{3}{2}},1) 
]\nonumber\\
&+&  \sin{{\textstyle\frac{\theta}{2}}}
[ \sqrt{{\textstyle\frac{2}{3}}}{\cal F}_{-1}^{1}
({\textstyle\frac{1}{2}},1)
+ \sqrt{{\textstyle\frac{3}{2}}}(1+\cos{\theta})
[{\cal F}_{1}^{1}({\textstyle\frac{3}{2}},1)
-{\textstyle\frac{1}{\sqrt{3}}}{\cal  
F}_{1}^{2}({\textstyle\frac{3}{2}},1)]
- {\textstyle\frac{1}{\sqrt{6}}}(3\cos{\theta}+1)\nonumber\\ & \times  
&
[{\cal F}_{-1}^{1}({\textstyle\frac{3}{2}},1)
-\sqrt{3}{\cal F}_{-1}^{2}({\textstyle\frac{3}{2}},1)]
] \\
H_{1{\textstyle\frac{1}{2}}; {\textstyle\frac{1}{2}}1} 
&=&\cos{{\textstyle\frac{\theta}{2}}}
[ 
-{\textstyle\frac{2}{\sqrt{3}}}{\cal  
F}_{-1}^{1}({\textstyle\frac{1}{2}},1)
+ {\textstyle\frac{\sqrt{3}}{2}}(1-\cos{\theta})
[{\cal F}_{1}^{1}({\textstyle\frac{3}{2}},1)
-{\textstyle\frac{1}{\sqrt{3}}}{\cal  
F}_{1}^{2}({\textstyle\frac{3}{2}},1)]
+ {\textstyle\frac{1}{2\sqrt{3}}}(3\cos{\theta}-1)\nonumber\\ & \times  
&
[{\cal F}_{-1}^{1}({\textstyle\frac{3}{2}},1)
-\sqrt{3}{\cal F}_{-1}^{2}({\textstyle\frac{3}{2}},1)]
]\nonumber\\
&+&  \sin{{\textstyle\frac{\theta}{2}}}
[ 
-\sqrt{{\textstyle\frac{2}{3}}}
[{\cal F}_{0}^{0}({\textstyle\frac{1}{2}},1)
-{\cal F}_{0}^{1}({\textstyle\frac{1}{2}},1)]
- {\textstyle\frac{1}{\sqrt{6}}}(3\cos{\theta}+1)
[{\cal F}_{0}^{1}({\textstyle\frac{3}{2}},1)
-{\cal F}_{0}^{2}({\textstyle\frac{3}{2}},1)]
- (1+\cos{\theta}) {\cal F}_{-2}^{2}({\textstyle\frac{3}{2}},1) ] 
\end{eqnarray}
\begin{eqnarray}
H_{-1-{\textstyle\frac{1}{2}};-{\textstyle\frac{1}{2}}1} 
&=&\cos{{\textstyle\frac{\theta}{2}}}
[ 
-{\textstyle\frac{2}{\sqrt{3}}}{\cal  
F}_{1}^{1}({\textstyle\frac{1}{2}},1)
+ {\textstyle\frac{\sqrt{3}}{2}}(1-\cos{\theta})
[{\cal F}_{-1}^{1}({\textstyle\frac{3}{2}},1)
+{\textstyle\frac{1}{\sqrt{3}}}{\cal  
F}_{-1}^{2}({\textstyle\frac{3}{2}},1)] 
+ {\textstyle\frac{1}{2\sqrt{3}}}(3\cos{\theta}-1)\nonumber\\ & \times  
&
[{\cal F}_{1}^{1}({\textstyle\frac{3}{2}},1)
+\sqrt{3}{\cal F}_{1}^{2}({\textstyle\frac{3}{2}},1)]
]\nonumber\\
&+&  \sin{{\textstyle\frac{\theta}{2}}}
[ 
-\sqrt{{\textstyle\frac{2}{3}}}
[{\cal F}_{0}^{0}({\textstyle\frac{1}{2}},1)+{\cal  
F}_{0}^{1}({\textstyle\frac{1}{2}},1)]
+ {\textstyle\frac{1}{\sqrt{6}}}(3\cos{\theta}+1)
[{\cal F}_{0}^{1}({\textstyle\frac{3}{2}},1)+{\cal  
F}_{0}^{2}({\textstyle\frac{3}{2}},1)]
- (1+\cos{\theta}) {\cal F}_{2}^{2}({\textstyle\frac{3}{2}},1) 
] 
\\
H_{0-{\textstyle\frac{1}{2}};-{\textstyle\frac{1}{2}}1} 
&=&\cos{{\textstyle\frac{\theta}{2}}}
[
{\textstyle\frac{1}{\sqrt{3}}}
[{\cal F}_{0}^{0}({\textstyle\frac{1}{2}},1)+{\cal  
F}_{0}^{1}({\textstyle\frac{1}{2}},1)]
- {\textstyle\frac{1}{\sqrt{3}}}(3\cos{\theta}-1)
[{\cal F}_{0}^{1}({\textstyle\frac{3}{2}},1)+{\cal  
F}_{0}^{2}({\textstyle\frac{3}{2}},1)]
-\sqrt{2} (1-\cos{\theta}) {\cal F}_{2}^{2}({\textstyle\frac{3}{2}},1) 
 ]\nonumber\\
&+&  \sin{{\textstyle\frac{\theta}{2}}}
[-\sqrt{{\textstyle\frac{2}{3}}}{\cal  
F}_{1}^{1}({\textstyle\frac{1}{2}},1)
- \sqrt{{\textstyle\frac{3}{2}}}(1+\cos{\theta})
[{\cal F}_{-1}^{1}({\textstyle\frac{3}{2}},1)
+{\textstyle\frac{1}{\sqrt{3}}}{\cal  
F}_{-1}^{2}({\textstyle\frac{3}{2}},1)] 
+ {\textstyle\frac{1}{\sqrt{6}}}(3\cos{\theta}+1)\nonumber\\ & \times  
&
[{\cal F}_{1}^{1}({\textstyle\frac{3}{2}},1)
+\sqrt{3}{\cal F}_{1}^{2}({\textstyle\frac{3}{2}},1)]
 ] 
\\
H_{1-{\textstyle\frac{1}{2}};-{\textstyle\frac{1}{2}}1} 
&=&\cos{{\textstyle\frac{\theta}{2}}}
[ 
{\textstyle\frac{\sqrt{3}}{2}}(1+\cos{\theta})
[{\cal F}_{-1}^{1}({\textstyle\frac{3}{2}},1)
+{\textstyle\frac{1}{\sqrt{3}}}{\cal  
F}_{-1}^{2}({\textstyle\frac{3}{2}},1)] 
+{\textstyle\frac{\sqrt{3}}{2}}(1-\cos{\theta})
[{\cal F}_{1}^{1}({\textstyle\frac{3}{2}},1)
+\sqrt{3}{\cal F}_{1}^{2}({\textstyle\frac{3}{2}},1)]
]\nonumber\\
&+&  \sin{{\textstyle\frac{\theta}{2}}}
[ -\sqrt{{\textstyle\frac{3}{2}}}(1+\cos{\theta})
[{\cal F}_{0}^{1}({\textstyle\frac{3}{2}},1)
+{\cal F}_{0}^{2}({\textstyle\frac{3}{2}},1)]
-(1-\cos{\theta}) {\cal F}_{2}^{2}({\textstyle\frac{3}{2}},1)] 
\end{eqnarray}
\begin{eqnarray}
H_{-1-{\textstyle\frac{1}{2}};{\textstyle\frac{1}{2}}1} 
&=&\cos{{\textstyle\frac{\theta}{2}}}
[ 
\sqrt{{\textstyle\frac{2}{3}}}
[{\cal F}_{0}^{0}({\textstyle\frac{1}{2}},1)
-{\cal F}_{0}^{1}({\textstyle\frac{1}{2}},1)]
- {\textstyle\frac{1}{\sqrt{6}}}(3\cos{\theta}-1)
[{\cal F}_{0}^{1}({\textstyle\frac{3}{2}},1)
-{\cal F}_{0}^{2}({\textstyle\frac{3}{2}},1)]
+ (1-\cos{\theta}) {\cal F}_{-2}^{2}({\textstyle\frac{3}{2}},1) 
]\nonumber\\
&+&  \sin{{\textstyle\frac{\theta}{2}}}
[ {-\textstyle\frac{2}{\sqrt{3}}}
{\cal F}_{-1}^{1}({\textstyle\frac{1}{2}},1)
+ {\textstyle\frac{\sqrt{3}}{2}}(1+\cos{\theta})
[{\cal F}_{1}^{1}({\textstyle\frac{3}{2}},1)
-{\textstyle\frac{1}{\sqrt{3}}}{\cal  
F}_{1}^{2}({\textstyle\frac{3}{2}},1)]
- {\textstyle\frac{1}{2\sqrt{3}}}(3\cos{\theta}+1)\nonumber\\ & \times  
&
[{\cal F}_{-1}^{1}({\textstyle\frac{3}{2}},1)
-\sqrt{3}{\cal F}_{-1}^{2}({\textstyle\frac{3}{2}},1)]]
\\
H_{0-{\textstyle\frac{1}{2}};{\textstyle\frac{1}{2}}1} 
&=&\cos{{\textstyle\frac{\theta}{2}}}
[ \sqrt{{\textstyle\frac{2}{3}}}{\cal  
F}_{-1}^{1}({\textstyle\frac{1}{2}},1)
+ \sqrt{{\textstyle\frac{3}{2}}}(1-\cos{\theta})
[{\cal F}_{1}^{1}({\textstyle\frac{3}{2}},1)
-{\textstyle\frac{1}{\sqrt{3}}}{\cal  
F}_{1}^{2}({\textstyle\frac{3}{2}},1)]
+ {\textstyle\frac{1}{\sqrt{6}}}(3\cos{\theta}-1)\nonumber\\ & \times  
&
[{\cal F}_{-1}^{1}({\textstyle\frac{3}{2}},1)
-\sqrt{3}{\cal F}_{-1}^{2}({\textstyle\frac{3}{2}},1)]]
\nonumber\\
&+&  \sin{{\textstyle\frac{\theta}{2}}}
[ 
{\textstyle\frac{1}{\sqrt{3}}}
[{\cal F}_{0}^{0}({\textstyle\frac{1}{2}},1)
-{\cal F}_{0}^{1}({\textstyle\frac{1}{2}},1)]
- {\textstyle\frac{1}{\sqrt{3}}}(3\cos{\theta}+1)
[{\cal F}_{0}^{1}({\textstyle\frac{3}{2}},1)
-{\cal F}_{0}^{2}({\textstyle\frac{3}{2}},1)]
- \sqrt{2}(1+\cos{\theta}) {\cal F}_{-2}^{2}
({\textstyle\frac{3}{2}},1) ]
\\
H_{1-{\textstyle\frac{1}{2}};{\textstyle\frac{1}{2}}1} 
&=&\cos{{\textstyle\frac{\theta}{2}}}[
-\sqrt{{\textstyle\frac{3}{2}}}(1-\cos{\theta})
[{\cal F}_{0}^{1}({\textstyle\frac{3}{2}},1)
-{\cal F}_{0}^{2}({\textstyle\frac{3}{2}},1)]
+ (1+\cos{\theta})
{\cal F}_{-2}^{2}({\textstyle\frac{3}{2}},1) 
]
\nonumber\\ 
&+&  \sin{{\textstyle\frac{\theta}{2}}}
[
{{\textstyle\frac{\sqrt{3}}{2}}}(1+\cos{\theta})
[{\cal F}_{-1}^{1}({\textstyle\frac{3}{2}},1)
-\sqrt{3}{\cal F}_{-1}^{2}({\textstyle\frac{3}{2}},1)]
+ {{\textstyle\frac{\sqrt{3}}{2}}}(1-\cos{\theta})
[{\cal F}_{1}^{1}({\textstyle\frac{3}{2}},1)
-{\textstyle\frac{1}{\sqrt{3}}}{\cal  
F}_{1}^{2}({\textstyle\frac{3}{2}},1)]
], 
\end{eqnarray}
\end{widetext}
which are identifiable respectively with  $H_{i,\mu_{s}}, 
i=1,...,4$ of \cite{tab961}. It may be noted that the above 
formulae are not restricted to threshold production but are 
applicable at higher energies as well. Even at low energies, the higher order 
partial waves could be expected to play a role in determining the spin observables.  

Likewise, in anticipation of future experimental developments, explicit 
expressions for the helicity amplitudes for 
photo production of higher spin mesons may 
also be obtained, not only at threshold but also at all energies  
using  \eqref{me} in \eqref{ha}. 
Our approach thus provides incidentally an elegant methodology to 
obtain the helicity amplitudes in terms of the partial wave 
multipole amplitudes for all orders not only in the case of vector meson 
photo production but also for photo production of mesons 
with arbitrary spin parity $s^\pi$, even if $s$ is greater than 1.

\end{document}